\documentclass[aps,prd,preprintnumbers,twocolumn,amsmath,amssymb,superscriptaddress,showpacs,floatfix]{revtex4}
\usepackage{graphicx}
\usepackage{dcolumn}
\usepackage{bm}

\begin{document}

\preprint{LBNL-57478}

\title{Modified Fragmentation Function from Quark Recombination}

\author{A. Majumder}
\affiliation{Nuclear Science Division, Lawrence Berkeley Laboratory,
Berkeley, California 94720}

\author{Enke Wang}
\affiliation{Institute of Particle Physics, Huazhong Normal University,
         Wuhan 430079, China}
\affiliation{Nuclear Science Division, Lawrence Berkeley Laboratory,
Berkeley, California 94720}

\author{Xin-Nian Wang}
\affiliation{Nuclear Science Division, Lawrence Berkeley Laboratory,
Berkeley, California 94720}

\date{May 1, 2005}

\begin{abstract}
Within the framework of the constituent quark model, it is shown that
the single hadron fragmentation function of a parton can be expressed as
a convolution of shower diquark or triquark distribution function
and quark recombination probability, if the interference between
amplitudes of quark recombination with different momenta is neglected.
The recombination probability is determined by the hadron's wavefunction
in the constituent quark model. The shower diquark or triquark distribution
functions of a fragmenting jet are defined in terms of overlapping
matrices of constituent quarks and parton field operators. They are
similar in form to dihadron or trihadron fragmentation functions in 
terms of parton operator and hadron states. Extending the formalism to 
the field theory at finite temperature, we automatically derive 
contributions to the effective single hadron fragmentation function 
from the recombination of shower and thermal constituent quarks. Such 
contributions involve single or diquark distribution functions which in 
turn can be related to diquark or triquark distribution functions via 
sum rules. We also derive QCD evolution equations for quark distribution 
functions that in turn determine the evolution of the effective jet 
fragmentation functions in a thermal medium.
\end{abstract}


\maketitle

\section{Introduction}
\label{sec1}

In the study of the properties of dense matter and search for quark gluon
plasma in high-energy heavy-ion collisions, jet 
quenching \cite{Gyulassy:1990ye} or the suppression of leading high $p_T$ 
hadrons \cite{Wang:1991xy}
has become a powerful diagnostic tool. The theoretical concept relies on
the study of parton propagation in medium and induced radiative energy
loss \cite{Gyulassy:1993hr,Baier:1994bd,Zakharov:1996fv,Wiedemann:2000tf}.
Using perturbative QCD (pQCD) and taking into account the
intricate Landau-Pomeranchuk-Migdal (LPM) \cite{lpm} interference effect,
one can calculate the radiative energy loss. It has a unique dependence 
on the path length traversed by the parton and the local gluon density of the 
medium. Such energy loss is manifested in the modification of the effective
parton fragmentation functions \cite{frg} which in turn leads to the
suppression of the single inclusive hadron spectra. In non-central
heavy-ion collisions, the path length dependence of the energy loss
gives rise to the azimuthal angle dependence of the high $p_T$ single 
hadron spectra \cite{v2}. Therefore, experimental measurements
of the suppression of single hadron spectra or the modification of the
fragmentation functions and their centrality dependence can provide
important information about the initial gluon density and geometry
of the produced dense matter. Moreover, one can go beyond single
hadron spectra and study the modification of multiple hadron
correlations inside jets due to multiple scattering of the
parton in the medium \cite{Majumder:2004pt}.

Data from experiments at the Relativistic Heavy-ion Collider (RHIC) have
indeed confirmed the predicted features of jet quenching \cite{Jacobs:2004qv}.
One has seen not only a significant suppression of single inclusive high $p_T$
hadron spectra  \cite{phenix1,star1} and strong azimuthal
angle dependence \cite{star-v2}, but also the suppression of
high $p_T$ hadrons on the opposite side of a triggered high $p_T$ hadron
in the central $Au+Au$ collisions \cite{star-jet}. These are all 
consistent with the qualitative features of jet quenching due to parton 
energy loss \cite{Wang:2003mm}. The extracted initial gluon density in
the most central $Au+Au$ collisions at $\sqrt{s}=200$ GeV is found to 
be about 30 times higher than that in a cold $Au$
nuclei \cite{Wang:2003mm,ww02,Vitev:2002pf}. Combined with the
enormous collective expansion as measured by the elliptic flow of the
final bulk hadrons \cite{Ackermann:2000tr}, current experimental data
points to the formation of a strongly interactive quark-gluon
plasma in central $Au+Au$ collisions at RHIC.

During the propagation and interaction inside a deconfined hot
partonic medium, a fast parton can have not only induced gluon radiation
but also induced absorption of the surrounding thermal gluons.
The detailed balance leads to an energy dependence of the net 
energy loss that is stronger than without for an intermediate energy 
parton \cite{Wang:2001cs}. In principle, one can consider gluon
absorption as a parton recombination process and it continues 
until the hadronization of the bulk partonic matter. Eventually, during
the hadronization, partons from the jet can combine with those
from the medium to form the final hadrons. Indeed, there already 
exists some evidence for such parton recombination in the measured 
hadron spectra in heavy-ion collisions at RHIC. At
intermediate $p_T=2-4$ GeV/$c$, the suppression of baryons
is significantly smaller than mesons, leading to a baryon to meson 
ratio larger than 1 \cite{Adler:2003cb}. This is about
a factor of 5 increase over the value in $p+p$ collisions.
On the other hand, the azimuthal anisotropy of the baryon spectra is
larger than that of meson spectra. Such a flavor dependence of the
nuclear modification of the hadron spectra and their azimuthal
anisotropy is not consistent with a picture of pure parton energy
loss followed by vacuum fragmentation. The most striking empirical 
observation of the flavor dependence, that could reveal the underlying 
hadronization mechanism, is the scaling behavior between the
azimuthal anisotropy of baryon and meson spectra \cite{Adams:2003am},
$v_2^M(p_T/2)/2=v_2^B(p_T/3)/3$. Such an observation is inspired by 
a schematic model of hadron  production by constituent quark 
recombination \cite{Voloshin:2002wa}.
Here, $v_2^M$($v_2^B$) is the second coefficient of the Fourier
transformation of the azimuthal angle distribution of mesons (baryons).

Many quark recombination models \cite{Hwa:2002tu,Greco:2003xt,Fries:2003vb}
are successful in describing the observed flavor dependence of the nuclear
modification of hadron spectra at intermediate $p_T$. These models in 
general involve thermal quarks in the medium and employ the constituent 
quark model for hadron wavefunctions which determine the recombination 
probabilities. They, however, differ in the handling of specific
recombination processes. Some considered only recombination
of thermal quarks, with inherent correlations caused by jet quenching.
Others also include recombination of thermal and shower quarks from the 
fragmenting jet, which dominate the hadron spectra at intermediate $p_T$. 
They also differ in the determination of the constituent quark
distributions from high $p_T$ jets and there exist ambiguities in the
connection between partons from pQCD hard processes and constituent
quarks that form the final hadrons. 

One of the models that we will follow closely in this paper is by 
Hwa and Yang \cite{Hwa:2002tu}.
In this model, quark recombination processes are traced back to parton
fragmentation processes in vacuum. They assume that the initial
produced hard partons will evolve into a shower of constituent
quarks which then recombine to form the final hadrons in the
hadronization process. The formulation of the recombination of the 
shower quarks of the partonic jets and the medium quarks in heavy-ion 
collisions is straightforward, given both the shower and medium
quark distributions. Since the Hwa-Yang model is a phenomenological
one, the nuclear modification of the shower quark distributions and
their QCD evolution cannot be calculated. The model
has to rely on fitting to the experimentally measured hadron spectra
to obtain the corresponding nuclear modified shower quark distributions
for each centrality of heavy-ion collisions and correlations between
shower quarks are completely neglected.

In this paper, we make a first attempt to derive the recombination
model of jet fragmentation functions from a field theoretical
formulation and the constituent quark model of hadron structure.
Within the constituent quark model, we consider the parton fragmentation
as a two stage process. The initial parton first evolves into a shower
of constituent quarks that subsequently will combine with each other
to form the final hadrons. Since constituent quarks are non-perturbative
objects in QCD just like hadrons, the conversion of hard partons into
showers of constituent quarks is not calculable in pQCD. However,
we can define constituent quark distributions in a jet as overlapping
matrices of the parton field operator and the constituent quark states,
similarly as one has defined hadron fragmentation functions.
Ignoring interferences in the process of quark recombination, we
demonstrate that the single inclusive meson (baryon) fragmentation
functions can be cast as a convolution of the diquark (triquark)
distribution functions and the recombination probabilities, which
are determined by the hadrons' wavefunctions. This is similar in
spirit to the Hwa-Yang recombination model. Given a form of the
hadrons' wavefunctions in the constituent quark model, one can
in principle extract constituent quark distribution functions
from measured jet fragmentation functions. We are also
able to derive the Dokshitzer-Gribov-Lipatov-Altarelli-Parisi (DGLAP)
\cite{Dokshitzer:1977sg,Gribov:1972ri,Altarelli:1977zs} evolution
equations for the diquark and triquark distribution functions,
which in turn give rise to the DGLAP evolution equations for
the hadron fragmentation functions within the quark recombination
formalism. Such a reformulation of the jet fragmentation functions 
does nothing to simplify the description of jet hadronization.
However, extending the formalism to the case of jet fragmentation 
in a thermal medium and working within field theory at finite 
temperature, we can automatically derive contributions
from recombination between shower and thermal quarks in addition
to soft hadron production from recombination of thermal quarks and
leading hadrons from recombination of shower quarks. The shower
and thermal quark recombination involves single (diquark) quark
distribution functions which can be obtained from diquark (triquark)
distributions through sum rules. Therefore, one can consistently
describe three different processes within this formalism. One can
also consistently take into account parton energy loss and detailed
balance effect for jet fragmentation inside a thermal medium.

For convenience in this paper, we will simply refer to the constituent
quarks as quarks and the initial current quarks and gluons as partons.
The quarks from jet fragmentation will be denoted as shower quarks in
contrast to quarks in the thermal medium in heavy-ion collisions. The
remainder of the paper is organized as follows: In Section II, we
review single inclusive hadron fragmentation functions in terms
of overlapping matrix elements between parton field operators and
hadronic states. In Section III and IV, we introduce the hadronic
wavefunctions in the constituent quark model and reformulate the
parton fragmentation
functions in terms of quark recombination probabilities and shower
quark distribution functions of the fragmenting partons, which
are defined as the overlapping matrices of the parton field operators
and constituent quark states. We also derive the QCD evolution
equations for the shower quark distribution functions and discuss
possible sum rules relating triquark, diquark and single quark distribution
functions. In Section V, we extend the formalism to parton fragmentation
in a thermal medium within the framework of field theory at finite
temperature and derive various contributions to the effective
fragmentation functions from shower-thermal and thermal-thermal
quark recombination. We summarize the results and discuss future
work in Section VI.

\begin{widetext}

\section{Single Hadron Fragmentation Function}
\label{sec2}

For hadron production processes, in $e^+e^-$ annihilation for example,
that involve a large momentum scale, the inclusive cross section at
leading twist can be factorized into a hard part for parton
scattering at short distances and a soft part for hadronization
at long distances as the parton fragmentation function. Though one 
can systematically calculate the hard part in pQCD due to asymptotic 
freedom, the parton fragmentation function is nonperturbative
and currently can only be measured in experiments. However, one
can derive the DGLAP evolution equations with the momentum scale, which 
have been tested successfully against the experimental data \cite{bethke}.
In this section, we review the parton fragmentation functions as
defined in the form of matrix elements of parton operators and the
corresponding DGLAP evolution equations.

As in a previous study of medium modification of fragmentation functions
with detailed balance \cite{Osborne:2002dx}, we consider $e^+e^-$ annihilation
to illustrate the parton operator definition of the quark fragmentation
functions. Within the collinear factorization approximation,
the inclusive differential cross section  can be expressed as
 \begin{eqnarray}
  \frac{d\sigma_{e^+e^-\rightarrow h}}{dz_{h}}
  =\frac{1}{2s}\frac{e^4}{q^4}
  L_{\mu\nu}(p_a,p_b)\frac{dW^{\mu\nu}}{dz_h}
=\sum_q\sigma_{0}^{q\bar q}
    \left[D_q^{h}(z_{h})
    +D_{\bar q}^{h}(z_{h})\right]\, ,
 \label{inclusec1}
 \end{eqnarray}
where $q=p_a+p_b$ is the four-momentum of virtual photon
and $s=q^2\equiv Q^2$ is the invariant mass of $e^+e^-$ pair. The leptonic
and hadronic tensors are given by
$L_{\mu\nu}(p_a,p_b)=(1/4){\rm Tr}(\gamma_{\mu}{\not\!p}_a
    \gamma_{\nu}{\not\!p}_b)$
and
 \begin{eqnarray}
  W^{\mu\nu}(q)=\sum_{X}
  \langle 0\left|J^{\mu}(0)\right|X\rangle\langle X
  \left|J^{\nu}(0)\right|0\rangle
  (2\pi)^4\delta^4(p_{_X}-q)
   =\int d^4x e^{-i q\cdot x}
   \langle 0\left|J^{\mu}(0)J^{\nu}(x)\right|0\rangle\, ,
 \label{hadron1}
 \end{eqnarray}
where $\sum_X$ runs over all possible intermediate states and the
quark electromagnetic current is $J_{\mu}=\sum_q e_q{\bar
\psi}_q\gamma_{\mu}\psi_q$.  Here, $e_q$ is the fractional charge
of the quark in units of an electron charge. The total cross-section
at the lowest order in pQCD is
$\sigma_{0}^{q\bar q}\equiv N_c 4\pi\alpha^2 e^2_q/3s$, where $N_c=3$ 
is the number of colors in the fundamental representation of SU(3) 
and $\alpha$ is the electromagnetic coupling constant. 
The single inclusive fragmentation functions of a current
quark and antiquark are defined as \cite{CSS89,CSS85,CQ89,MW04},
 \begin{eqnarray}
  D_{q}^{h}(z_{h})
  &=&\frac{z_{h}^3}{2}\int\frac{d^4p}{(2\pi)^4}
   \delta\left(z_{h} {-} \frac{p_{h}\cdot n}{p\cdot n}\right)
   \int d^4 xe^{-ip\cdot x}
   {\rm Tr}\left\lbrack\frac{\gamma\cdot n}{2 p_{h}\cdot n}
    \sum_S\langle 0\left|{\psi}(0)\right| S,p_{h}\rangle
    \langle p_{h},S\left|\overline\psi(x)\right|0\rangle\right\rbrack\, ,
 \label{qfrag1}\\
  D_{\bar q}^{h}(z_{h})
  &=&\frac{z_{h}^3}{2}\int\frac{d^4p}{(2\pi)^4}
   \delta\left(z_{h} {-} \frac{p_{h}\cdot n}{p\cdot n}\right)
   \int d^4 xe^{-ip\cdot x}
   {\rm Tr}\left\lbrack
    \sum_S\langle 0\left|{\overline\psi}(0)\right| S,p_{h}\rangle
    \frac{\gamma\cdot n}{2 p_{h}\cdot n}
    \langle p_{h}, S \left|\psi(x)\right|0\rangle\right\rbrack\, ,
 \label{qfrag2}
 \end{eqnarray}
respectively, where $p_{h}$ is the four-momentum of the identified
hadron and $z_{h}$ is its momentum fraction with respect to the initial
parton momentum. Sum over the color index
of the quark field is implicitly implied.
The light-like vector is defined as $n^\mu=[n^+, n^-,
n_{\perp}] =[0, 1,0_\perp]$. These fragmentation functions
$D_{q({\bar q})}^{h}(z_{h})$ can be interpreted as the
multiplicity distribution of hadrons with fractional momenta
between $z_{h}$ and $z_{h}+dz_{h}$ produced in the fragmentation of the
initial parton quark $q$ or antiquark $\bar q$.

 \begin{figure}
\resizebox{2.5in}{2.0in}{\includegraphics{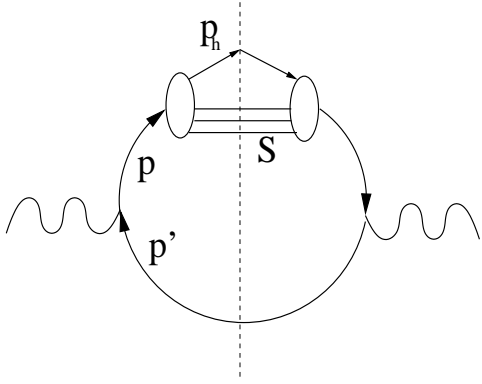}}
 \caption{diagram for single hadron fragmentation in $e^+e^-$
   annihilation.}
 \label{fig1}
 \end{figure}

In deriving Eqs.~(\ref{inclusec1}), (\ref{qfrag1}) and (\ref{qfrag2})
from Eq.~(\ref{hadron1}), the intermediate states
$\sum_X \left|X\rangle\langle X\right|$ have been replaced by two
complete subsets
$\sum_{S,S'} \left|S,p_h;S' \rangle\langle S'; S,p_h\right|$,
corresponding to independent fragmentation of the quark and
anti-quark created from the $e^+e^-$ annihilation.
One can further replace $|S' \rangle$ by a complete set
consisting of quarks and gluons. To the lowest order of pQCD,
only a single quark or anti-quark state $|p' \rangle$
of $|S' \rangle$ contributes to the hadronic tensor
in Eq.~(\ref{hadron1}). Thus, one can effectively express,
 \begin{equation}
  \sum_{X}\left|X\rangle\langle X\right|=
  \sum_{S}\int\frac{d^3p_{h}}{(2\pi)^3 2E_{h}}
  \int\frac{d^3 p'}{(2\pi)^3 2E_{p'}}|S, p_{h};p'\rangle
  \langle p';p_{h},S|\, ,
 \label{XX1}
 \end{equation}
and the factorized form of the differential cross section
in Eq.~(\ref{inclusec1}) can be represented by a cut diagram
as shown in Fig.~\ref{fig1}, in which the amplitudes of the
fragmentation processes are represented by the blob
connecting initial quark and final hadron states. This is normally
presented as cut-vertices \cite{Osborne:2002dx}.
Since the hadronic states $|X\rangle$ are
all color singlet, the above approximation implies that $|S\rangle$
also carries color which can only be neutralized via exchange of
soft partons. These soft processes are higher twist and are
suppressed by powers of $1/Q^2$. Soft gluon exchanges also lead
to eikonal contributions or gauge links that ensure the gauge
invariance of the defined quark fragmentation functions.
They, however, do not appear in the light-cone gauge, $n\cdot A=0$.

Similarly, the hadron fragmentation function $D_g^{h}(z_{h})$ from
a gluon parton is defined as
 \begin{eqnarray}
  D_g^{h}(z_{h})
  &=&\frac{z_{h}^2}{2}\int\frac{d^4p}{(2\pi)^4}
   \delta\left(z_{h}-\frac{p_{h}\cdot n}{p\cdot n}\right)
   \int d^4 xe^{-ip\cdot x}
   d_{\mu\nu}(p)\sum_{S}\langle 0\left|A^{\mu}(0)
    \right|S, p_{h}\rangle
    \langle p_{h},S\left|A^{\nu}(x)\right|0\rangle\, ,
 \label{gfrag1}
 \end{eqnarray}
where gluonic color indices are also implicitly summed over.
The gluon polarization tensor is defined in terms of the
polarization vectors $\varepsilon^\mu(k)$ and has the form
 \begin{eqnarray}
   d_{\mu\nu}(p)&=&\sum_{\lambda=1,2}
   \varepsilon_{\mu}(p,\lambda)\varepsilon_{\nu}(p,\lambda)
 \nonumber\\
   &=&-g_{\mu\nu}+\frac{p_{\mu}n_{\nu}+p_{\nu}n_{\mu}}{n\cdot p}\, ,
 \label{polar}
 \end{eqnarray}
in the light-cone gauge.

The scale dependence of the single inclusive hadron fragmentation
functions can be calculated by considering the radiative corrections to the
fragmentation functions at the next-to-leading order in
perturbation QCD \cite{CSS85,CQ89}. It is governed by the DGLAP
evolution  equations \cite{Dokshitzer:1977sg,Gribov:1972ri,Altarelli:1977zs},
 \begin{eqnarray}
   Q^2\,\frac{d}{d Q^2}D_q^{h}(z_{h}, Q^2)
   &=&\frac{\alpha_s(Q^2)}{2\pi}\int_{z_{h}}^1\frac{dz}{z}
   \left[\gamma_{qq}(z)
   D_q^{h}(\frac{z_{h}}{z},Q^2)
   +\gamma_{qg}(z)
   D_g^{h}(\frac{z_{h}}{z}, Q^2)\right]\, ,
 \label{eq:DGLAP1}\\
   Q^2\,\frac{d }{d Q^2}D_g^{h}(z_{h}, Q^2)
   &=&\frac{\alpha_s(Q^2)}{2\pi}\int_{z_{h}}^1\frac{dz}{z}
   \left[\gamma_{gq}(z)
   D_s^{h}(\frac{z_{h}}{z}, Q^2)
   +\gamma_{gg}(z)
   D_g^{h}(\frac{z_{h}}{z},Q^2)\right]\, ,
 \label{eq:DGLAP2}
 \end{eqnarray}
where $D_s^{h}(z_{h}/z, Q^2)$ is the singlet quark fragmentation
function
 \begin{equation}
  D_s^{h}(z,Q^2)\equiv\sum_{q}\left[
  D_q^{h}(z,Q^2) +D_{\bar q}^{h}(z,Q^2)\right]\, ,
 \label{singlet1}
 \end{equation}
and the splitting functions are \cite{Field}
 \begin{mathletters}
 \label{split}
 \begin{eqnarray}
  \gamma_{qq}(z)&=& C_F\Bigl[\frac{1+z^2}{(1-z)_+}
    +\frac{3}{2}\delta(1-z)\Bigr]\, ,
 \label{split1}\\
   \gamma_{qg}(z)&=& C_F\frac{1+(1-z)^2}{z}\, ,
 \label{split2}\\
  \gamma_{gq}(z)&=&T_F
    \Bigl[ z^2+(1-z)^2\Bigr]\, ,
 \label{split3}\\
   \gamma_{gg}(z)&=& 2C_A\Bigl[\frac{z}{(1-z)_+}
     +\frac{1-z}{z}+z(1-z)\Bigr]
     +\delta(1-z)\Bigl[\frac{11}{6}C_A-\frac{2}{3}n_f T_F\Bigr]\, .
 \label{split4}
 \end{eqnarray}
 \end{mathletters}
Here, $n_f$ is the number of quark flavors, the SU($N_c$) Casimirs
are given by $C_F=(N_c^2-1)/2N_c$, $C_A=N_c$ and $T_F=1/2$. The
`+'-function is defined such that the replacement
 \begin{equation}
 \label{+function}
  \int_0^1 dz\frac{f(z)}{(1-z)_+}=\int_0^1 dz
   \frac{f(z)-f(1)}{1-z}\,
 \end{equation}
is valid for any function $f(z)$ that is continuous at $z=1$.

The single hadron fragmentation functions as defined in Eqs.~(\ref{qfrag1}),
(\ref{qfrag2}) and (\ref{gfrag1}) satisfy the following momentum
sum rules,
 \begin{equation}
 \sum_{h}\int dz z D_a^{h}(z,Q^2)=1, (a=q,\bar q, g).
 \label{sum11}
 \end{equation}
One can also naively define the zeroth moments of the fragmentation
functions,
 \begin{eqnarray}
   \sum_{h}\int dz D_a^{h}(z,Q^2)
     &=&{\overline N_q^h}\, ,
 \label{sum12}\\
\end{eqnarray}
which are simply the average hadron multiplicities of the parton
jets. In principle, the average hadron multiplicities are
not infrared safe and therefore not well defined in the simple
leading log and leading twist approximation.
One has to go beyond these approximations
and take into account the coherence of parton cascade. In practice,
one can introduce a cut-off to regularize the infrared behavior.

\section{Quark Recombination and Parton Fragmentation}
\label{sec3}

\subsection{The Constituent Quark Model}
\label{sec30}

In hard processes involving a large momentum scale, the produced
partons are normally off-shell and subject to radiative processes
following the hard scattering. The DGLAP evolution equations describe
these radiative processes for a highly virtual parton in pQCD.
Eventually, however, when the parton's virtuality becomes smaller
than allowed for the applicability of pQCD, further interaction among 
produced or shower partons and the ensuing hadronization can never
be described by pQCD. Therefore, one cannot calculate perturbatively
the parton fragmentation functions with momentum scale $Q<Q_0$.
Conventionally, one uses experimental data to parameterize the
fragmentation functions with  $Q\leq Q_0$ (called initial condition)
and the DGLAP equations [Eqs.~(\ref{eq:DGLAP1}) and (\ref{eq:DGLAP2})]
predict the evolution of the fragmentation functions with the
momentum scale.

In this paper, we will employ a constituent quark model to
describe the soft interaction and hadronization of shower
partons. In this approach, soft interaction between shower
partons below the scale $Q_0$ will be effectively represented
by constituent quarks and their interaction. 
The non-perturbative conversion between
shower partons and constituent quarks will be described by the
constituent quark distribution functions in the fragmenting jet.
Further interaction among constituent quarks during the hadronization
will be given by the hadrons' wave functions in the constituent quark
model.

In this constituent quark model, a hadronic state
with momentum $p_h=[p_h^+,0,0_\perp]$ can be expressed as
 \begin{eqnarray}
  | p_h\rangle &=&\int
  [d^2k_{\perp}][dx]
  \varphi_h(k_{1\perp}, x_1; k_{2\perp}, x_2)
  |k_{1\perp}, x_1; k_{2\perp}, x_2\rangle ,
\label{pm}
\end{eqnarray}
for a meson and
\begin{eqnarray}
  | p_h\rangle &=&\int[d^2k_{\perp}][dx]
  \varphi_h(k_{1\perp}, x_1; k_{2\perp}, x_2; k_{3\perp}, x_3)
  |k_{1\perp}, x_1; k_{2\perp}, x_2; k_{3\perp}, x_3\rangle\, ,
 \label{pb}
 \end{eqnarray}
for a baryon. Here,
 \begin{eqnarray}
  [d^2k_{\perp}] &\equiv & 2(2\pi)^3\delta^{(2)}
  \left(\sum_{i=1}^n k_{i\perp}\right)
  \prod_{i=1}^n\frac{d^2 k_{i\perp}}{2(2\pi)^3}\, ,\quad
  \text{( $n=2$ for a meson and 3 for a baryon)}\, ,
  \label{d2k}\\
  {[dx]} &\equiv & \delta\left(1-\sum_{i=1}^n x_i\right)
  \prod_{i=1}^n \frac{dx_i}{\sqrt{x_i}}\, , \qquad x_i=\frac{k_i^+}{p_h^+}\, ,
  \label{dx}
 \end{eqnarray}
and $\varphi_h $ is the hadronic wavefunction. In the infinite momentum frame,
the constituent quarks can be considered approximately on shell.
The normalization of the hadronic wavefunction is
\begin{eqnarray}
  \int [d^2k_{\perp}][dx]
  \left|\varphi_h(k_{i\perp},x_i)\right|^2=1\, ,
  \label{norm}
 \end{eqnarray}
given the normalization of single constituent quark states,
\begin{eqnarray}
  \langle k_{i\perp},x_i|k_{j\perp},x_j\rangle
  =(2\pi)^3 2x_i \delta^{(2)}\left(k_{i\perp}-k_{j\perp}\right)
  \delta\left(x_i-x_j\right)\, .
  \label{inner}
 \end{eqnarray}

\subsection{Double Constituent Quark Distribution Functions}
\label{sec3a}

Substituting the hadronic state in Eq.(\ref{pm}) in the constituent
quark model into Eq.~(\ref{qfrag1}),
we can express the meson fragmentation as
 \begin{eqnarray}
  D_{q}^{M}(z_M)
  &=&\frac{z_{M}^3}{2}\int\frac{d^4p}{(2\pi)^4}
   \delta\left(z_{M} {-} \frac{p_M^+}{p^+}\right)
   \int d^4 xe^{-ip\cdot x}
   [d^2k_{\perp}][dx]
   [d^2k_{\perp}'][dx']
 \nonumber\\
   &&\varphi_M(k_{1\perp}, x_1; -k_{1\perp}, 1-x_1)
   \varphi^*_M(k'_{1\perp}, x'_1; -k'_{1\perp}, 1-x'_1)
 \nonumber\\
   &&{\rm Tr}\left\lbrack\frac{\gamma^+}{2 p_{M}^+}
    \sum_S\langle 0\left|{\psi}(0)
    \right|S; k_{1\perp}, x_1; -k_{1\perp}, 1-x_1\rangle
   \langle -k'_{1\perp}, 1-x'_1; k'_{1\perp}, x'_1; S
   \left|\overline\psi(x)\right|0\rangle\right\rbrack .
\label{eq:frg-m1}
 \end{eqnarray}

 \begin{figure}
\resizebox{2.5in}{1.5in}{\includegraphics{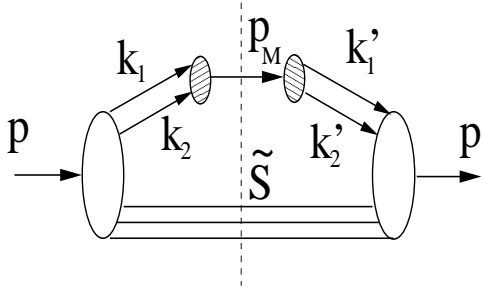}}
 \caption{The cut diagram for single meson fragmentation functions in terms
   of quark recombination. The blob connecting initial quark and final
   constituent quarks before recombination represents the amplitudes
   of shower-to-constituent quark conversion of the fragmenting quark. 
   Hadronic wavefunction is presented by the shaded blob.}
 \label{fig2}
 \end{figure}

Since the constituent quarks can be regarded as a complete set of
intermediate states $|\widetilde S \rangle$, they are equivalent to and 
can replace the complete set of hadronic states $|S \rangle$ in the 
above expression,
$\sum_{S}|S \rangle \langle S| =\sum_{\widetilde S}|\widetilde S \rangle
\langle \widetilde S|$. One can then interpret the above expression
in terms of quark recombination as shown in Fig.~\ref{fig2}:
a quark and anti-quark with
momentum $[x_1p^+_M, k_{1\perp}]$ and $[(1-x_1)p_M^+, -k_{1\perp}]$
from the fragmentation of the parton jet will combine to form a meson
with momentum $p_M=[p^+_M, 0, 0_{\perp}]$.
Since the fragmentation function $D_{q}^{M}(z_M)$ is proportional to
the single inclusive cross section for hadron production, it should naturally
include the interference between recombination of quark and anti-quark
pairs with different momenta that sum to the meson's momentum.
With the presence of the interference contribution, one can never
arrive at a probabilistic interpretation as postulated in all the
current recombination models \cite{Hwa:2002tu,Greco:2003xt,Fries:2003vb}.
However, if the hadronic
wavefunction $\varphi_M(k_{1\perp}, x_1; k_{2\perp}, x_2)$ in the
constituent quark model is sharply peaked, one may neglect
the interference contributions. Effectively,
one can complete the integration $\int [d^2k_{\perp}'][dx']$
in Eq.~(\ref{eq:frg-m1}) and assume that the final result is proportional
to the diagonal term with a coefficient $C_M$. One has then approximately,
 \begin{eqnarray}
   D_{q}^{M}(z_M) &\approx& C_M\frac{z_{M}^3}{2}\int\frac{d^4p}{(2\pi)^4}
   \delta\left(z_{M} {-} \frac{p_{M}^+}{p^+}\right)
   \int d^4 xe^{-ip\cdot x}
   \int\frac{d^2 k_{1\perp}}{(2\pi)^3}
   \frac{dx_1}{4x_1(1-x_1)}
   |\varphi_M(k_{1\perp}, x_1; -k_{1\perp}, 1-x_1)|^2
 \nonumber\\
   &&{\rm Tr}\left\lbrack\frac{\gamma^+}{2 p_M^+}
    \sum_{\widetilde S}\langle 0\left|{\psi}(0)
    \right|\widetilde S; k_{1\perp}, x_1; -k_{1\perp}, 1-x_1\rangle
   \langle -k_{1\perp}, 1-x_1; k_{1\perp}, x_1; \widetilde S
   \left|\overline\psi(x)\right|0\rangle\right\rbrack\, ,
 \label{qfragm01}
 \end{eqnarray}
where $C_M$ is a constant with dimensions of momentum that accounts for
the non-diagonal (or interference) contributions. Note that
 \begin{eqnarray}
  x_i=\frac{k_i^+}{p_M^+}=\frac{k_i^+}{p^+}\frac{p^+}{p_M^+}
  =\frac{z_i}{z_M}\, ,
  \label{x}
 \end{eqnarray}
and $z_i$ is the fractional momentum carried by the constituent quark.
Defining the diquark recombination probability $R_M$ as
 \begin{equation}
R_M(k_{1\perp},\frac{z_1}{z_M})
   \equiv
 \left|\varphi_M (k_{1\perp}, \frac{z_1}{z_M}; -k_{1\perp},
   1-\frac{z_1}{z_M})\right|^2 \, ,
 \label{Rm}
 \end{equation}
we can approximate the meson fragmentation function [Eq.~(\ref{qfragm01})] as
 \begin{eqnarray}
  D_{q}^{M}(z_M)
  \approx C_M\int_0^{z_M}\frac{dz_1}{2}
  R_M(0_\perp,\frac{z_1}{z_M})
  F_{q}^{q_1{\bar q}_2}(z_1, z_M-z_1)  \, ,
 \label{qfragm02}
 \end{eqnarray}
where the double  constituent quark (or diquark) distribution function in
a quark jet is defined as
 \begin{eqnarray}
  F_{q}^{q_1{\bar q}_2}(z_1,z_2)
  &=&\frac{z_{M}^4}{2z_1z_2}
  \int^{\Lambda}\frac{d^2 k_{1\perp}}{2(2\pi)^3}
  \int\frac{d^4p}{(2\pi)^4}
   \delta\left(z_M - \frac{p_M^+}{p^+}\right)
   \int d^4 xe^{-ip\cdot x}
  \nonumber\\
   &&{\rm Tr}\left\lbrack\frac{\gamma^+}{2 p_{M}^+}
    \sum_{\widetilde S}\langle 0\left|{\psi}(0)\right|\widetilde S, k_1,k_2\rangle
    \langle k_2,k_1,\widetilde S\left|\overline\psi(x)
    \right|0\rangle\right\rbrack\, .
 \label{qfrag03}
 \end{eqnarray}
Here, $p_{M}=k_1+k_2$, $z_M=z_1+z_2$, and $\Lambda$ is the cutoff for
the intrinsic transverse momentum of the constituent quark inside a
hadron, as provided by the hadron wavefunction. One can consider
$\Lambda$ as the scale for hadronization. One can also similarly
express single inclusive hadron fragmentation functions for antiquarks
and gluons in terms of the same diquark recombination
probability $R_M(x_1,x_2)$
and double constituent quark distribution functions from antiquark
and gluon jets, $F_{\bar q}^{q_1 \bar q_2}(z_1,z_2)$
and $F_g^{q_1 \bar q_2}(z_1,z_2)$, respectively, which are defined as
 \begin{eqnarray}
  F_{\bar q}^{q_1 \bar q_2}(z_1,z_2)
  &=&\frac{z_{M}^4}{2z_1z_2}
  \int^{\Lambda}\frac{d^2 k_{1\perp}}{2(2\pi)^3}
  \int\frac{d^4p}{(2\pi)^4}
   \delta\left(z_{M} {-} \frac{p_{M}^+}{p^+}\right)
   \int d^4 xe^{-ip\cdot x}
  \nonumber\\
   &&{\rm Tr}\left\lbrack
    \sum_{\widetilde S}\langle 0\left|{\overline\psi}(0)\right|\widetilde S, k_1,k_2\rangle
    \frac{\gamma^+}{2 p_{M}^+}
    \langle  k_2,k_1,\widetilde S\left|\psi(x)\right|0\rangle\right\rbrack\, ,
 \label{qfrag04}\\
  F_g^{q_1 \bar q_2}(z_1,z_2)
  &=&\frac{z_{M}^3}{2z_1 z_2}
  \int^{\Lambda}\frac{d^2 k_{1\perp}}{2(2\pi)^3}
  \int\frac{d^4p}{(2\pi)^4}
   \delta\left(z_{M}-\frac{p_{M}^+}{p^+}\right)
   \int d^4 xe^{-ip\cdot x}
  \nonumber\\
   && d_{\mu\nu}(p)\sum_{\widetilde S}\langle 0\left|A^{\mu}(0)
    \right|\widetilde S, k_1,k_2\rangle
    \langle k_2,k_1,\widetilde S\left|A^{\nu}(x)\right|0\rangle\, .
 \label{gfrag02}
 \end{eqnarray}
As we will show in the next subsection, the above definitions of diquark
distribution functions have almost the same form as dihadron quark
fragmentation functions \cite{MW04}, except the cutoff $\Lambda$ for
the intrinsic transverse momentum which is similar to the cut-off
$\mu_\perp$ for the intra-jet transverse momentum in dihadron
fragmentation functions.

The quark recombination process in our constituent quark model happens
at a scale $\mu\sim \Lambda$ during the evolution of the jet.
Therefore, the expression for hadron fragmentation function,
Eq. (\ref{qfragm02}), in this model is valid for jets produced at
any initial scale $Q$. Since the recombination happens at a fixed
momentum scale $\mu$, the recombination probability should be independent
of the initial scale $Q$. The scale dependence of the hadron fragmentation
functions $D_q^M(z_M,Q^2)$ will be completely determined by the scale
dependence of the diquark distribution 
functions $F_q^{q_1\bar{q_2}}(z_1,z_2,Q^2)$.
Similar to the derivation of the
evolution equations of the dihadron fragmentation function \cite{MW04},
one can also obtain the radiative corrections to the double
constituent quark distributions from parton fragmentation,
 \begin{eqnarray}
   F_{q}^{q_1\bar q_2}(z_1,z_2,Q^2)
   &=&F_{q}^{q_1\bar q_2}(z_1,z_2,\mu^2)+\frac{\alpha_s(\mu^2)}{2\pi}
   \int_{\mu^2}^{Q^2}\frac{dk_{\perp}^2}{k_{\perp}^2}
   \int_{z_1+z_2}^1\frac{dz}{z^2}\gamma_{qq}(z)
   F_q^{q_1\bar q_2}\left(\frac{z_1}{z},\frac{z_2}{z},\mu^2\right)
 \nonumber\\
   &&+\frac{\alpha_s(\mu^2)}{2\pi}
   \int_{\mu^2}^{Q^2}\frac{dk_{\perp}^2}{k_{\perp}^2}
   \int_{z_1+z_2}^1\frac{dz}{z^2}\gamma_{qg}(z)
   F_g^{q_1\bar q_2}\left(\frac{z_1}{z},\frac{z_2}{z},
   \mu^2\right)
 \nonumber\\
   &&+\frac{\alpha_s(\mu^2)}{2\pi}
   \int_{\mu^2}^{\Lambda^2}\frac{dl_{\perp}^2}{l_{\perp}^2}
   \sum_{i=1}^2\int_{z_i}^{1-z_{\bar i}}
   \frac{dz}{z(1-z)}{\hat \gamma}_{qq}(z)
   F_q^{q_i}\left(\frac{z_i}{z},\mu^2\right)
   F_g^{\bar q_{\bar i}}\left(\frac{z_{\bar i}}{1-z},\mu^2\right)\, ,
 \label{rad21}\\
   F_{g}^{q_1\bar q_2}(z_1,z_2,Q^2)
   &=&F_{g}^{q_1\bar q_2}(z_1,z_2,\mu^2)+\frac{\alpha_s(\mu^2)}{2\pi}
   \int_{\mu^2}^{Q^2}\frac{dk_{\perp}^2}{k_{\perp}^2}
   \int_{z_1+z_2}^1\frac{dz}{z^2}\gamma_{gq}(z)
   F_s^{q_1\bar q_2}\left(\frac{z_1}{z},\frac{z_2}{z},\mu^2\right)
 \nonumber\\
   &&+\frac{\alpha_s(\mu^2)}{2\pi}
   \int_{\mu^2}^{Q^2}\frac{dk_{\perp}^2}{k_{\perp}^2}
   \int_{z_1+z_2}^1\frac{dz}{z^2}\gamma_{gg}(z)
   F_g^{q_1\bar q_2}\left(\frac{z_1}{z},\frac{z_2}{z}, \mu^2\right)
 \nonumber\\
  &&+\frac{\alpha_s(\mu^2)}{2\pi}
   \int_{\mu^2}^{\Lambda^2}\frac{dl_{\perp}^2}{l_{\perp}^2}
   \sum_{i=1}^2\sum_q\int_{z_i}^{1-z_{\bar i}}\frac{dz}{z(1-z)}\gamma_{gq}(z)
   F_q^{q_i}\left(\frac{z_i}{z},\mu^2\right)
   F_{\bar q}^{\bar q_{\bar i}}\left(\frac{z_{\bar i}}{1-z},\mu^2\right)
 \nonumber\\
   &&+\frac{\alpha_s(\mu^2)}{2\pi}
   \int_{\mu^2}^{\Lambda^2}\frac{dl_{\perp}^2}{l_{\perp}^2}
   \int_{z_1}^{1-z_2}\frac{dz}{z(1-z)}{\hat \gamma}_{gg}(z)
   F_g^{q_1}\left(\frac{z_1}{z},\mu^2\right)
   F_g^{\bar q_2}\left(\frac{z_2}{1-z},\mu^2\right)\, .
 \label{rad22}
 \end{eqnarray}
The singlet constituent diquark distribution
function is defined as
 \begin{equation}
  F_s^{q_1\bar q_2}(z_1, z_2,\mu^2)\equiv\sum_{q}\left[
  F_q^{q_1 \bar q_2}(z_1, z_2,\mu^2)
  +F_{\bar q}^{q_1 \bar q_2}(z_1, z_2,\mu^2)\right]\, .
 \label{singlet2}
 \end{equation}
The splitting $\gamma$-functions are the same as
in the evolution equation of single inclusive hadron fragmentation
functions given in Eqs.(\ref{split1})-(\ref{split4}). The first two
terms in Eqs.~(\ref{rad21}) and (\ref{rad22}) represent diquark
production from a single parton after the radiative splitting. The third term
in Eq.~(\ref{rad21}) and the third and fourth terms in Eq.~(\ref{rad22})
are from independent single quark production from each of the partons
after the radiative splitting. The indices ${\bar i}=2, 1$
and $\bar q_{\bar i}=\bar q_2, q_1$
for $i=1, 2$ and $q_i=q_1, \bar q_2$,  representing the exchange
between $q_1$ and $\bar q_2$ in the independent single quark production
process. These terms from independent single quark
production are proportional to the products of two single constituent
quark distribution functions,
 \begin{eqnarray}
  F_{q}^{q}(z_{q})
  &=&\frac{z_{q}^3}{2}\int\frac{d^4p}{(2\pi)^4}
   \delta\left(z_q - \frac{p_q^+}{p^+}\right)
   \int d^4 xe^{-ip\cdot x}
   {\rm Tr}\left\lbrack\frac{\gamma\cdot n}{2 p_q^+}
    \sum_{\widetilde S}\langle 0\left|{\psi}(0)\right|\widetilde S,p_{q}\rangle
    \langle p_{q},\widetilde S\left|\overline\psi(x)\right|0\rangle\right\rbrack\, ,
 \label{qdistr1}\\
  F_{\bar q}^{q}(z_{q})
  &=&\frac{z_{q}^3}{2}\int\frac{d^4p}{(2\pi)^4}
   \delta\left(z_{q} - \frac{p_{q}^+}{p^+}\right)
   \int d^4 xe^{-ip\cdot x}
   {\rm Tr}\left\lbrack
    \sum_{\widetilde S}\langle 0\left|{\overline\psi}(0)\right|\widetilde S,p_{q}\rangle
    \frac{\gamma\cdot n}{2 p_{q}^+}
    \langle p_{q}, \widetilde S \left|\psi(x)\right|0\rangle\right\rbrack\, ,
 \label{qbardistr} \\
  F_g^{q}(z_{q})
  &=&\frac{z_{q}^2}{2}\int\frac{d^4p}{(2\pi)^4}
   \delta\left(z_{q}-\frac{p_{q}^+}{p^+}\right)
   \int d^4 xe^{-ip\cdot x}
   d_{\mu\nu}(p)\sum_{\widetilde S}\langle 0\left|A^{\mu}(0)
    \right|\widetilde S, p_{q}\rangle
    \langle p_{q},\widetilde S\left|A^{\nu}(x)\right|0\rangle\, ,
 \label{gdistr}
 \end{eqnarray}
whose definitions have the same form as single hadron fragmentation
functions in Eqs.~(\ref{qfrag1}),(\ref{qfrag2}) and (\ref{gfrag1}),
replacing the single hadron state $h$ with a single constituent quark $q$.

In the independent single quark production, there are no
virtual corrections. Therefore, the corresponding
splitting ${\hat \gamma}$-functions
\begin{eqnarray}
  {\hat \gamma}_{qq}(z)&=& C_F\frac{1+z^2}{1-z}\, ,
 \label{split5}\\
  {\hat \gamma}_{gg}(z)&=&2C_A\left[\frac{z}{1-z}
     +\frac{1-z}{z}+z(1-z)\right]\, ,
 \label{split6}
\end{eqnarray}
have no '$+$'-function and delta functions, unlike $\gamma_{qq}$
and $\gamma_{gg}$ in Eqs.(\ref{split1})-(\ref{split4}).

Since the hadrons' wavefunctions in the constituent quark model
restrict the relative transverse momentum of the constituent quarks
within a hadron to a finite value,
we have an intrinsic transverse momentum cutoff $\Lambda$ in the
definition of the double constituent quark distributions that are
relevant for quark recombination in Eqs.~(\ref{qfrag03})--(\ref{gfrag02}).
Similarly, such a restriction should also be
applied to the relative transverse momentum between the two
quarks from the independent single quark production processes
in the radiative corrections. This is why there is a cutoff $\Lambda$
in the relative transverse momentum in the third term in
Eq.~(\ref{rad21}) and the third and four terms in Eq.~(\ref{rad22})
for the independent production of two constituent quarks that will
form a final hadron. This cutoff is determined by the hadronic structure
in the constituent quark model and is independent of the momentum
scale of the initial hard parton scattering. Therefore,
these radiative corrections do not
contribute to the DGLAP evolution equations for the diquark
distribution functions. Differentiating Eqs.~(\ref{rad21}) and
(\ref{rad22}) with respect to $Q^2$, one obtains,
 \begin{mathletters}
 \label{DGLAPq1}
 \begin{eqnarray}
   Q^2\,\frac{d}{d Q^2}F_q^{q_1 \bar q_2}(z_1, z_2, Q^2)
   &=&\frac{\alpha_s(Q^2)}{2\pi}\int_{z_1+z_2}^1\frac{dz}{z^2}
   \left[\gamma_{qq}(z)
   F_q^{q_1 \bar q_2}(\frac{z_1}{z},\frac{z_2}{z},Q^2)
   +\gamma_{qg}(z)
   F_g^{q_1 \bar q_2}(\frac{z_1}{z},\frac{z_2}{z}, Q^2)\right]\, ,
 \label{DGLAPq11}\\
   Q^2\,\frac{d }{d Q^2}F_g^{q_1 \bar q_2}(z_1, z_2, Q^2)
   &=&\frac{\alpha_s(Q^2)}{2\pi}\int_{z_1+z_2}^1\frac{dz}{z^2}
   \left[\gamma_{gq}(z)
   F_s^{q_1 \bar q_2}(\frac{z_1}{z},\frac{z_2}{z}, Q^2)
   +\gamma_{gg}(z)
   F_g^{q_1 \bar q_2}(\frac{z_1}{z},\frac{z_2}{z},Q^2)\right]\, .
 \label{DGLAPq12}
 \end{eqnarray}
 \end{mathletters}
One can show that combining the above evolution equations
for the double constituent quark distribution functions with
the expression of jet fragmentation function in Eq.~(\ref{qfragm02}),
the DGLAP evolution equations for single inclusive hadron
fragmentation functions in Eqs.~(\ref{eq:DGLAP1}) and (\ref{eq:DGLAP2})
can be recovered.

Similar to the single and dihadron fragmentation
functions, one can also derive sum rules for the
single and diquark distribution functions in a parton jet. One
can define the mean constituent quark multiplicities from the single
quark distributions as
\begin{equation}
\int_0^1 dz F_a^q(z)\equiv {\overline N_a^q}, (a=q,\bar q, g).
\label{s11}
\end{equation}
Similarly, single quark distribution functions also obey
the momentum sum rule,
\begin{equation}
\int_0^1 dz \sum_q zF_a^q(z)=1 \, .
\label{s12}
\end{equation}
For diquark distribution functions
$F_{a}^{q_1,q_2}(z_1, z_2,Q^2)$,
the multiplicity sum rule leads to the second
cumulant moments of the multiplicity distribution,
 \begin{eqnarray}
   \int dz_1 dz_2
   F_a^{q_1 q_2}(z_1, z_2,Q^2)
     &\equiv &{\overline {N_a^{q_1}(N_a^{q_2}-\delta_{q_1 q_2})}}\, ,
 \label{s22}
\end{eqnarray}
where $\delta_{q_1 q_2}=1$ if $q_1$ and $q_2$ are identical quarks
and $\delta_{q_1 q_2}=0$ if $q_1$ and $q_2$ are different.
It illustrates the importance of quark correlations in the diquark
distribution functions. Any initial conditions for the diquark
distributions should contain the correlation between two quarks
within a jet. For
two identical quark distributions, the above equation also illustrates
the information on the multiplicity fluctuation contained in the diquark
distributions. Because of the correlation and fluctuation contained in the
diquark distributions, one cannot find any rigorous sum rules relating
single and diquark distributions. Following the work on dihadron
fragmentation functions \cite{MW04}, an approximate ansatz that contains
the minimum correlation and fluctuation is
\begin{equation}
   \sum_{q_2}\int dz_2
   F_{a}^{q_1 q_2}(z_1, z_2,Q^2)
     \approx\frac{\overline {N_{a}^{q_1}(N_{a}^{q_2}-\delta_{q_1,q_2})}}
     {\overline N_{a}^{q_1}}F_{a}^{q_1}(z_1, Q^2)\, .
 \label{s23}
 \end{equation}
Using the momentum sum rule Eq.~(\ref{s12}) for single quark
distributions, the above ansatz will lead to the momentum sum rule
for double constituent quark distributions,
 \begin{equation}
   \sum_{q_1q_2}\int dz_1dz_2 \frac{1}{2}(z_1+z_2)
   F_{a}^{q_1 q_2}(z_1, z_2,Q^2)
     =\frac{\overline {N_{a}^{q_1}(N_{a}^{q_2}-\delta_{q_1,q_2})}}
     {2\overline {N_{a}}^{q_1}\, \overline {N_{a}}^{q_2} }
(\overline {N_{a}}^{q_1}+ \overline {N_{a}}^{q_2}) \, .
 \label{s26}
 \end{equation}
Whether such ansatz is a good approximation remains to be explored.

\subsection{Dihadron Fragmentation Functions}
\label{sec3b}

The definitions of diquark distribution functions of a jet are very
similar to that of dihadron fragmentation functions, except that the
constituent quarks in the distributions are restricted by the hadron
wavefunction in the phase space for quark recombination during
hadronization. According to Ref.~\cite{MW04}, the dihadron fragmentation 
functions of a quark or gluon parton are defined as,
 \begin{eqnarray}
  D_{q}^{h_1 h_2}(z_{h_1},z_{h_2})
  &=&\frac{z_{h}^4}{2z_{h_1}z_{h_2}}
  \int\frac{d^2 p_{h_{1\perp}}}{2(2\pi)^3}
  \int\frac{d^4p}{(2\pi)^4}
   \delta\left(z_h - \frac{p_h^+}{p^+}\right)
   \int d^4 xe^{-ip\cdot x}
  \nonumber\\
   &&{\rm Tr}\left\lbrack\frac{\gamma^+}{2 p_{h}^+}
    \sum_{S}\langle 0\left|{\psi}(0)\right|S, p_{h_1},p_{h_2}\rangle
    \langle p_{h_2},p_{h_1},S\left|\overline\psi(x)\right|0\rangle\right\rbrack\, ,
 \label{qfrag3}\\
  D_{\bar q}^{h_1 h_2}(z_{h_1},z_{h_2})
  &=&\frac{z_{h}^4}{2z_{h_1}z_{h_2}}
  \int\frac{d^2 p_{h_{1\perp}}}{2(2\pi)^3}
  \int\frac{d^4p}{(2\pi)^4}
   \delta\left(z_h - \frac{p_h^+}{p^+}\right)
   \int d^4 xe^{-ip\cdot x}
  \nonumber\\
   &&{\rm Tr}\left\lbrack
    \sum_{S}\langle 0\left|{\overline\psi}(0)\right|S, p_{h_1},p_{h_2}\rangle
    \frac{\gamma^+}{2 p_{h}^+}
    \langle  p_{h_2},p_{h_1},S\left|\psi(x)\right|0\rangle\right\rbrack\, ,
 \label{qfrag4} \\
  D_g^{h_1 h_2}(z_{h_1},z_{h_2})
  &=&\frac{z_{h}^3}{2z_{h_1}z_{h_2}}
  \int\frac{d^2 p_{h_1\perp}}{2(2\pi)^3}
  \int\frac{d^4p}{(2\pi)^4}
   \delta\left(z_{h}-\frac{p_{h}^+}{p^+}\right)
   \int d^4 xe^{-ip\cdot x}
  \nonumber\\
   && d_{\mu\nu}(p)\sum_{S}\langle 0\left|A^{\mu}(0)
    \right|S, p_{h_1},p_{h_2}\rangle
    \langle p_{h_2},p_{h_1},S\left|A^{\nu}(x)\right|0\rangle\, ,
 \label{gfrag2}
 \end{eqnarray}
where $p_{h}=p_{h_1}+p_{h_2}$ and $z_{h}=z_{h_1}+z_{h_2}$.
Note that the transverse momentum, $p_{h_1\perp}$ and $p_{h_2\perp}$,
are defined as perpendicular to the summed total momentum $p_{h}$.
The relative transverse momentum $q_T$ used in Ref.~\cite{MW04}
is then $\vec{q}_\perp=2\vec{p}_{h_1\perp}=-2\vec{p}_{h_2\perp}$.

The radiative corrections to the dihadron fragmentation functions
are similar to that of diquark distribution functions, except that
the transverse momentum between two individual hadrons is not limited
as in the case for diquark distributions due to quark recombination
during hadronization. Therefore, the cutoff $\Lambda$ in Eq.~(\ref{rad21})
and (\ref{rad22}) should be replaced by $Q^2$. The DGLAP evolution
equations for dihadron fragmentation functions then become \cite{MW04},
 \begin{eqnarray}
   Q^2\frac{d}{d Q^2}
   D_q^{h_1 h_2}(z_{h_1}, z_{h_2}, Q^2)
 &=&\frac{\alpha_s(Q^2)}{2\pi}
   \left[\int_{z_{h_1}+z_{h_2}}^1\frac{dz}{z^2}\gamma_{qq}(z)
   D_q^{h_1 h_2}\left(\frac{z_{h_1}}{z},\frac{z_{h_2}}{z},Q^2\right)\right.
 \nonumber\\
  &&+\sum_{i=1}^2\int_{z_{h_i}}^{1-z_{h_{\bar i}}}\frac{dz}{z(1-z)}{\hat \gamma}_{qq}(z)
   D_q^{h_i}\left(\frac{z_{h_i}}{z},Q^2\right)
   D_g^{h_{\bar i}}\left(\frac{z_{h_{\bar i}}}{1-z},Q^2\right)
 \nonumber\\
   &&+\left.\int_{z_{h_1}+z_{h_2}}^1\frac{dz}{z^2}\gamma_{qg}(z)
   D_g^{h_1 h_2}\left(\frac{z_{h_1}}{z},\frac{z_{h_2}}{z},
   Q^2\right)\right]\, ,
 \label{DGLAP21}\\
   Q^2\frac{d}{d Q^2}
   D_g^{h_1 h_2}(z_{h_1}, z_{h_2}, Q^2)
 &=&\frac{\alpha_s(Q^2)}{2\pi}
   \left[\int_{z_{h_1}+z_{h_2}}^1\frac{dz}{z^2}\gamma_{gq}(z)
   D_s^{h_1 h_2}\left(\frac{z_{h_1}}{z},\frac{z_{h_2}}{z},Q^2\right)\right.
 \nonumber\\
  &&+\sum_{i=1}^2\sum_{q}\int_{z_{h_i}}^{1-z_{h_{\bar i}}}\frac{dz}{z(1-z)}\gamma_{gq}(z)
   D_q^{h_i}\left(\frac{z_{h_i}}{z},Q^2\right)
   D_{\bar q}^{h_{\bar i}}\left(\frac{z_{h_{\bar i}}}{1-z},Q^2\right)
 \nonumber\\
   &&+\int_{z_{h_1}+z_{h_2}}^1\frac{dz}{z^2}\gamma_{gg}(z)
   D_g^{h_1 h_2}\left(\frac{z_{h_1}}{z},\frac{z_{h_2}}{z},
   Q^2\right)
 \nonumber\\
   &&+\left.\int_{z_{h_1}}^{1-z_{h_2}}\frac{dz}{z(1-z)}{\hat \gamma}_{gg}(z)
   D_g^{h_1}\left(\frac{z_{h_1}}{z},Q^2\right)
   D_g^{h_2}\left(\frac{z_{h_2}}{1-z},Q^2\right)\right]\, .
 \label{DGLAP22}
 \end{eqnarray}
The additional terms in comparison to the DGLAP evolution equations
for single hadron fragmentation functions that are proportional to
the convolution of two single hadron fragmentation functions arise
from independent fragmentation of both the leading parton and radiated
gluon. They can be attributed to the dependence of the relative 
transverse momentum between two hadrons from the independent 
fragmentation on the momentum scale. This is the main difference between the
evolution of the above dihadron fragmentation functions and the
diquark distributions in the quark recombination model.

\section{Baryon Fragmentation Functions in Quark Recombination Model}
\label{sec4}

\subsection{Triple Constituent Quark Distribution Function}
\label{sec4a}

Since a baryon has three constituent quarks, the fragmentation
functions for a parton into baryons naturally involve triple
quark distribution functions in the quark recombination model.
Similar to the meson fragmentation functions, we can also
express the baryon fragmentation functions in terms of baryon
wave functions and overlapping matrix elements of parton
fields and constituent quark states, as illustrated in Fig.~\ref{fig3},
 \begin{eqnarray}
  D_{q}^{B}(z_B)
  &=&\frac{z_{B}^3}{2}\int\frac{d^4p}{(2\pi)^4}
   \delta\left(z_{B} {-} \frac{p_{B}^+}{p^+}\right)
   \int d^4 xe^{-ip\cdot x} [d^2k_{\perp}][dx] [d^2k_{\perp}'][dx']
 \nonumber\\
   && \varphi_B(k_{1\perp}, x_1; k_{2\perp}, x_2; k_{3\perp}, x_3)
   \varphi^*_B(k'_{1\perp}, x'_1; k'_{2\perp}, x'_2; k'_{3\perp}, x'_3)
 \nonumber\\
   &&{\rm Tr}\left\lbrack\frac{\gamma^+}{2 p_{B}^+}
    \sum_{\widetilde S}\langle 0\left|{\psi}(0)
    \right|\widetilde S; k_{1\perp}, x_1; k_{2\perp}, x_2; k_{3\perp}, x_3\rangle
   \langle k'_{3\perp}, x'_3; k'_{2\perp}, x'_2; k'_{1\perp}, x'_1; \widetilde S
   \left|\overline\psi(x)\right|0\rangle\right\rbrack
 \nonumber\\
   &\approx& C_B\frac{z_{B}^3}{2}\int\frac{d^4p}{(2\pi)^4}
   \delta\left(z_{B} {-} \frac{p_{B}^+}{p^+}\right)
   \int d^4 xe^{-ip\cdot x}
\int\frac{d^2 k_{1\perp}}{(2\pi)^3}\frac{dx_1}{4x_1}
   \int\frac{d^2 k_{2\perp}}{(2\pi)^3}\frac{dx_2}{4x_2}\frac{1}{x_3}
   |\varphi_b(k_{1\perp}, x_1; k_{2\perp}, x_2; k_{3\perp}, x_3)|^2
 \nonumber\\
   &&{\rm Tr}\left\lbrack\frac{\gamma^+}{2 p_{B}^+}
    \sum_{\widetilde S}\langle 0\left|{\psi}(0)
    \right|\widetilde S; k_{1\perp}, x_1; k_{2\perp}, x_2; k_{3\perp}, x_3\rangle
   \langle k_{3\perp}, x_3; k_{2\perp}, x_2; k_{1\perp}, x_1; \widetilde S
   \left|\overline\psi(x)\right|0\rangle\right\rbrack\, ,
 \label{qfragm04}
 \end{eqnarray}
where $[d^2k_{\perp}]$ and $[dx]$ are given by Eqs.~(\ref{d2k}) and
(\ref{dx}) for baryons, $x_3=1-x_1-x_2$,
$k_{3\perp}=-k_{1\perp}-k_{2\perp}$,
and  $C_B$ is a constant with the dimension of momentum.
Notice again that
 \begin{equation}
  x_1=\frac{k_1^+}{p_B^+}=\frac{k_1^+}{p^+}\frac{p^+}{p_B^+}
  =\frac{z_1}{z_B}; \, \,
 x_2=\frac{k_2^+}{p_B^+}=\frac{k_2^+}{p^+}\frac{p^+}{p_B^+}
  =\frac{z_2}{z_B}.
  \label{x13}
 \end{equation}
As in the quark recombination model for meson fragmentation functions, we
also define the recombination probability for a baryon as
 \begin{equation}
   R_B(k_{1\perp},\frac{z_1}{z_B}; k_{2\perp},\frac{z_2}{z_B})
   \equiv |\varphi_B(k_{1\perp},\frac{z_1}{z_B} ; k_{2\perp},
   \frac{z_2}{z_B}; -k_{1\perp}-k_{2\perp}, 1-\frac{z_1+z_2}{z_B})|^2
  \, .
 \label{Rb}
 \end{equation}
The baryon fragmentation function from Eq.~(\ref{qfragm04}) can then be
cast in the form
 \begin{eqnarray}
  D_{q}^{B}(z_B)
  \approx {C_B}\int_0^{z_B}\frac{dz_1}{2}\int_0^{z_B-z_1}\frac{dz_2}{2}
  R_B(0_\perp,\frac{z_1}{z_B}; 0_\perp,\frac{z_2}{z_B})
  F_{q}^{q_1q_2q_3}(z_1, z_2, 1-z_1-z_2) \, .
 \label{qfragm05}
 \end{eqnarray}
This relation is very similar to that for meson fragmentation
functions in Eq.~(\ref{qfragm02}).
The triple constituent quark distribution function is defined
as
 \begin{eqnarray}
  F_{q}^{q_1q_2q_3}(z_1,z_2,z_3)
  &=&\frac{z_{B}^4}{2z_1z_2z_3}
  \int^{\Lambda}\frac{d^2 k_{1\perp}}{2(2\pi)^3}
  \int^{\Lambda}\frac{d^2 k_{2\perp}}{2(2\pi)^3}
  \int\frac{d^4p}{(2\pi)^4}
   \delta\left(z_{B} {-} \frac{p_{B}^+}{p^+}\right)
   \int d^4 xe^{-ip\cdot x}
  \nonumber\\
   &&{\rm Tr}\left\lbrack\frac{\gamma^+}{2 p_{B}^+}
    \sum_{\widetilde S}\langle 0\left|{\psi}(0)\right|\widetilde S, k_1,k_2,k_3\rangle
    \langle k_3,k_2,k_1,\widetilde S\left|\overline\psi(x)
    \right|0\rangle\right\rbrack\, ,
 \label{qfrag06}
 \end{eqnarray}
where, $p_{B}=k_1+k_2+k_3$, $z_B=z_1+z_2+z_3$ and $\Lambda$ is the
cutoff for the intrinsic transverse momentum of the constituent
quarks in a baryon. Similarly, the triple constituent quark distribution
function from the antiquark and gluon parton are defined as
 \begin{eqnarray}
  D_{\bar q}^{q_1 q_2 q_3}(z_1,z_2,z_3)
  &=&\frac{z_{B}^4}{2z_1z_2z_3}
   \int^{\Lambda}\frac{d^2 k_{1\perp}}{2(2\pi)^3}
   \int^{\Lambda}\frac{d^2 k_{2\perp}}{2(2\pi)^3}
   \int\frac{d^4p}{(2\pi)^4}
   \delta\left(z_{B} {-} \frac{p_{B}^+}{p^+}\right)
   \int d^4 xe^{-ip\cdot x}
  \nonumber\\
    &&{\rm Tr}\left\lbrack
    \sum_{\widetilde S}\langle 0\left|{\overline\psi}(0)\right|\widetilde S, k_1,k_2,k_3\rangle
    \frac{\gamma^+}{2 p_{B}^+}
    \langle k_3,k_2,k_1,\widetilde S\left|\psi(x)\right|0\rangle\right\rbrack\, ,
 \label{qfrag07}\\
   D_g^{q_1 q_2 q_3}(z_1,z_2,z_3)
   &=&\frac{z_{B}^3}{2z_1z_2z_3}
   \int^{\Lambda}\frac{d^2 k_{1\perp}}{2(2\pi)^3}
   \int^{\Lambda}\frac{d^2 k_{2\perp}}{2(2\pi)^3}
   \int\frac{d^4p}{(2\pi)^4}
   \delta\left(z_{B}-\frac{p_{B}^+}{p^+}\right)
   \int d^4 xe^{-ip\cdot x}
  \nonumber\\
   && d_{\mu\nu}(p)\sum_{\widetilde S}\langle 0\left|A^{\mu}(0)
    \right|\widetilde S,k_1,k_2,k_3\rangle
    \langle k_3,k_2,k_1,\widetilde S\left|A^{\nu}(x)\right|0\rangle\, .
 \label{gfrag03}
 \end{eqnarray}

 \begin{figure}
\resizebox{2.5in}{1.5in}{\includegraphics{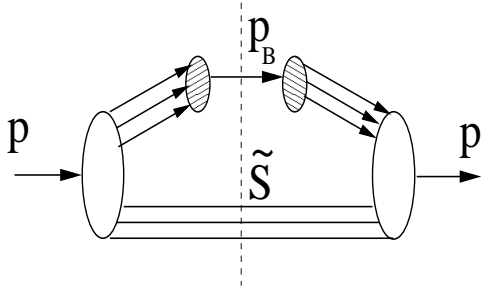}}
 \caption{The cut diagram for single baryon fragmentation function in terms
   of quark recombination.}
 \label{fig3}
 \end{figure}

Similarly as for the diquark distribution functions, one can also obtain
the radiative corrections to the triple constituent quark distribution
functions,
 \begin{eqnarray}
   &&F_{q}^{q_1q_2q_3}(z_1,z_2,z_3,Q^2)
   =F_{q}^{q_1q_2q_3}(z_1,z_2,z_3,\mu^2)
 \nonumber\\
  &&\qquad+\frac{\alpha_s(\mu^2)}{2\pi}
   \int_{\mu^2}^{Q^2}\frac{dk_{\perp}^2}{k_{\perp}^2}
   \int_{z_1+z_2+z_3}^1\frac{dz}{z^3}\gamma_{qq}(z)
   F_q^{q_1q_2q_3}\left(\frac{z_1}{z},\frac{z_2}{z},\frac{z_3}{z},\mu^2\right)
 \nonumber\\
   &&\qquad+\frac{\alpha_s(\mu^2)}{2\pi}
   \int_{\mu^2}^{Q^2}\frac{dk_{\perp}^2}{k_{\perp}^2}
   \int_{z_1+z_2+z_3}^1\frac{dz}{z^3}\gamma_{qg}(z)
   F_g^{q_1q_2q_3}\left(\frac{z_1}{z},\frac{z_2}{z},\frac{z_3}{z},
   \mu^2\right)
 \nonumber\\
   &&\qquad+\frac{\alpha_s(\mu^2)}{2\pi}
   \int_{\mu^2}^{\Lambda^2}\frac{dl_{\perp}^2}{l_{\perp}^2}
   \sum_{i=1}^3\int_{z_i}^{1-z_{i^\prime}-z_{i^{\prime\prime}}}
   \frac{dz}{z(1-z)^2}{\hat \gamma}_{qq}(z)
   F_q^{q_i}\left(\frac{z_i}{z},\mu^2\right)
   F_g^{q_{i^{\prime}}q_{i^{\prime\prime}}}\left(\frac{z_{i^{\prime}}}{1-z},
   \frac{z_{i^{\prime\prime}}}{1-z},\mu^2\right)
 \nonumber\\
   &&\qquad+\frac{\alpha_s(\mu^2)}{2\pi}
   \int_{\mu^2}^{\Lambda^2}\frac{dl_{\perp}^2}{l_{\perp}^2}
   \sum_{i=1}^3\int_{z_{i^\prime}+z_{i^{\prime\prime}}}^{1-z_i}
   \frac{dz}{z^2(1-z)}{\hat \gamma}_{qq}(z)
   F_q^{q_{i^{\prime}}q_{i^{\prime\prime}}}\left(\frac{z_{i^{\prime}}}{z},
   \frac{z_{i^{\prime\prime}}}{z},\mu^2\right)
   F_g^{q_i}\left(\frac{z_i}{1-z},\mu^2\right)\, ,
 \label{rad31}\\
   &&F_{g}^{q_1q_2q_3}(z_1,z_2,z_3,Q^2)
   =F_{g}^{q_1q_2q_3}(z_1,z_2,z_3,\mu^2)
 \nonumber\\
   &&\qquad+\frac{\alpha_s(\mu^2)}{2\pi}
   \int_{\mu^2}^{Q^2}\frac{dk_{\perp}^2}{k_{\perp}^2}
   \int_{z_1+z_2+z_3}^1\frac{dz}{z^3}\gamma_{gq}(z)
   F_s^{q_1q_2q_3}\left(\frac{z_1}{z},\frac{z_2}{z},\frac{z_3}{z},\mu^2\right)
 \nonumber\\
   &&\qquad+\frac{\alpha_s(\mu^2)}{2\pi}
   \int_{\mu^2}^{Q^2}\frac{dk_{\perp}^2}{k_{\perp}^2}
   \int_{z_1+z_2+z_3}^1\frac{dz}{z^2}\gamma_{gg}(z)
   F_g^{q_1q_2q_3}\left(\frac{z_1}{z},\frac{z_2}{z},\frac{z_3}{z}, \mu^2\right)
 \nonumber\\
  &&\qquad+\frac{\alpha_s(\mu^2)}{2\pi}
   \int_{\mu^2}^{\Lambda^2}\frac{dl_{\perp}^2}{l_{\perp}^2}
   \sum_{i=1}^3\sum_q\int_{z_i}^{1-z_{i^{\prime}}-z_{i^{\prime\prime}}}
   \frac{dz}{z(1-z)^2}\gamma_{gq}(z)
   F_q^{q_i}\left(\frac{z_i}{z},\mu^2\right)
   F_{\bar q}^{q_{i^\prime}q_{i^{\prime\prime}}}\left(\frac{z_{i^\prime}}{1-z},
   \frac{z_{i^{\prime\prime}}}{1-z},\mu^2\right)
 \nonumber\\
   &&\qquad+\frac{\alpha_s(\mu^2)}{2\pi}
   \int_{\mu^2}^{\Lambda^2}\frac{dl_{\perp}^2}{l_{\perp}^2}
   \sum_{i=1}^3\sum_q\int_{z_{i^\prime}+z_{i^{\prime\prime}}}^{1-z_i}
   \frac{dz}{z^2(1-z)}\gamma_{gq}(z)
   F_q^{q_{i^\prime}q_{i^{\prime\prime}}}\left(\frac{z_{i^\prime}}{z},
   \frac{z_{i^{\prime\prime}}}{z},\mu^2\right)
   F_{\bar q}^{q_i}\left(\frac{z_i}{1-z},\mu^2\right)
 \nonumber\\
   &&\qquad+\frac{\alpha_s(\mu^2)}{2\pi}
   \int_{\mu^2}^{\Lambda^2}\frac{dl_{\perp}^2}{l_{\perp}^2}
   \sum_{i=1}^3\int_{z_i}^{1-z_{i^\prime}-z_{i^{\prime\prime}}}
   \frac{dz}{z(1-z)^2}{\hat \gamma}_{gg}(z)
   F_g^{q_i}\left(\frac{z_i}{z},\mu^2\right)
   F_g^{q_{i^\prime}q_{i^{\prime\prime}}}\left(\frac{z_{i^\prime}}{1-z},
   \frac{z_{i^{\prime\prime}}}{1-z},\mu^2\right)\, ,
 \label{rad32}
 \end{eqnarray}
where $(i^\prime,i^{\prime\prime})=(2,3),(1,3),(1,2)$ as $i=1, 2, 3$
and the singlet triple constituent quark distribution is given by
 \begin{equation}
  F_s^{q_1q_2q_3}(z_1, z_2,z_3,\mu^2)\equiv\sum_{q}\left[
  F_q^{q_1q_2q_3}(z_1, z_2,z_3,\mu^2)
  +F_{\bar q}^{q_1q_2q_3}(z_1, z_2,z_3,\mu^2)\right]\, .
 \label{singlet3}
 \end{equation}

The DGLAP evolution equations for triple constituent quark distribution
functions can be obtained from the above,
 \begin{eqnarray}
   Q^2\,\frac{d}{d Q^2}D_q^{q_1 q_2 q_3}(z_1, z_2, z_3, Q^2)
   &=&\frac{\alpha_s(Q^2)}{2\pi}\int_{z_1+z_2+z_3}^1\frac{dz}{z^3}
   \left[\gamma_{qq}(z)
   D_q^{q_1 q_2 q_3}(\frac{z_1}{z},\frac{z_2}{z},\frac{z_3}{z},Q^2)\right.
 \nonumber\\
   &&\qquad+\left.\gamma_{qg}(z)
   D_g^{q_1 q_2 q_3}(\frac{z_1}{z},\frac{z_2}{z},\frac{z_3}{z},Q^2)\right]\, ,
 \label{DGLAPq21}\\
   Q^2\,\frac{d }{d Q^2}D_g^{q_1 q_2 q_3}(z_1, z_2, z_3, Q^2)
   &=&\frac{\alpha_s(Q^2)}{2\pi}\int_{z_1+z_2+z_3}^1\frac{dz}{z^3}
   \left[\gamma_{gq}(z)
   D_s^{q_1 q_2 q_3}(\frac{z_1}{z},\frac{z_2}{z},\frac{z_3}{z},Q^2)\right.
 \nonumber\\
   &&\qquad+\left.\gamma_{gg}(z)
   D_g^{q_1 q_2 q_3}(\frac{z_1}{z},\frac{z_2}{z},\frac{z_3}{z},Q^2)\right]\, .
 \label{DGLAPq22}
 \end{eqnarray}
Again, the radiative corrections from the independent fragmentation of
the leading and radiated partons do not contribute to the
DGLAP evolution because of the restriction on the relative
transverse momentum between constituent quarks by the baryon's wavefunction
during quark recombination.
As in the case for diquark distribution functions and meson fragmentation
functions, the above DGLAP evolution equations for triple quark
distributions will give rise to the DGLAP equations for baryon
fragmentation functions within the quark recombination model as
given by Eq.~(\ref{qfragm05}).

One can also obtain the multiplicity sum rule for the triple quark
distribution functions of a parton jet,
 \begin{equation}
   \int dz_1 dz_2 dz_3
   F_a^{q_1 q_2 q_3}(z_1, z_2,z_3,Q^2)
     ={\overline {N_a^{q_1}(N_a^{q_2}-\delta_{q_1q_2})
         (N_a^{q_3}-\delta_{q_2q_3}-\delta_{q_1q_3})}}\, .
 \label{sum31}
 \end{equation}
Similar to the ansatz of the sum rule relating single and diquark
distributions, one can also have the following ansatz for
a sum rule relating diquark and triple quark distributions,
 \begin{equation}
   \sum_{q_3}\int dz_3
   F_a^{q_1 q2 q_3}(z_1, z_2,z_3,Q^2)
     \approx \frac{\overline {N_a^{q_1}(N_a^{q_2}-\delta_{q_1q_2})
         (N_a^{q_3}-\delta_{q_2q_3}-\delta_{q_1q_3})}}
     {\overline {N_a^{q_1}(N_a^{q_2}-\delta_{q_1q_2})}}
     F_a^{q_1 q_2}(z_1, z_2,Q^2)\, .
 \label{sum33}
 \end{equation}

\subsection{Trihadron Fragmentation Function}
\label{sec4b}

It is straightforward to generalize the result for the triple quark
distribution functions to the trihadron fragmentation functions.
The operator expressions of the trihadron fragmentation
functions from quark, antiquark or a gluon jet are defined as
 \begin{eqnarray}
  D_{q}^{h_1 h_2 h_3}(z_{h_1},z_{h_2},z_{h_3})
  &=&\frac{z_{h}^4}{2z_{h_1}z_{h_2}z_{h_3}}
  \int\frac{d^2 p_{h_{1\perp}}}{2(2\pi)^3}
  \int\frac{d^2 p_{h_{2\perp}}}{2(2\pi)^3}
  \int\frac{d^4p}{(2\pi)^4}
   \delta\left(z_{h} {-} \frac{p_{h}^+}{p^+}\right)
   \int d^4 xe^{-ip\cdot x}
  \nonumber\\
   &&{\rm Tr}\left\lbrack\frac{\gamma^+}{2 p_{h}^+}
    \sum_{S}\langle 0\left|{\psi}(0)\right|S, p_{h_1},p_{h_2},p_{h_3}\rangle
    \langle p_{h_3},p_{h_2},p_{h_1},S
    \left|\overline\psi(x)\right|0\rangle\right\rbrack\, ,
 \label{qfrag5}\\
  D_{\bar q}^{h_1 h_2 h_3}(z_{h_1},z_{h_2},z_{h_3})
  &=&\frac{z_{h}^4}{2z_{h_1}z_{h_2}z_{h_3}}
  \int\frac{d^2 p_{h_{1\perp}}}{2(2\pi)^3}
  \int\frac{d^2 p_{h_{2\perp}}}{2(2\pi)^3}
  \int\frac{d^4p}{(2\pi)^4}
   \delta\left(z_{h} {-} \frac{p_{h}^+}{p^+}\right)
   \int d^4 xe^{-ip\cdot x}
  \nonumber\\
   &&{\rm Tr}\left\lbrack
    \sum_{S}\langle 0\left|{\overline\psi}(0)\right|S, p_{h_1},p_{h_2},p_{h_3}\rangle
    \frac{\gamma^+}{2 p_{h}^+}
    \langle p_{h_3},p_{h_2},p_{h_1},S\left|\psi(x)\right|0\rangle\right\rbrack\, ,
 \label{qfrag6} \\
  D_g^{h_1 h_2 h_3}(z_{h_1},z_{h_2},z_{h_3})
  &=&\frac{z_{h}^3}{2z_{h_1}z_{h_2}z_{h_3}}
  \int\frac{d^2 p_{h_{1\perp}}}{2(2\pi)^3}
  \int\frac{d^2 p_{h_{2\perp}}}{2(2\pi)^3}
  \int\frac{d^4p}{(2\pi)^4}
  \delta\left(z_{h}-\frac{p_{h}^+}{p^+}\right)
   \int d^4 xe^{-ip\cdot x}
  \nonumber\\
   && d_{\mu\nu}(p)\sum_{S}\langle 0\left|A^{\mu}(0)
    \right|S, p_{h_1},p_{h_2},p_{h_3}\rangle
    \langle p_{h_3},p_{h_2},p_{h_1},S\left|A^{\nu}(x)\right|0\rangle\, ,
 \label{gfrag3}
 \end{eqnarray}
where $p_{h}=p_{h_1}+p_{h_2}+p_{h_3}$ and
$z_{h}=z_{h_1}+z_{h_2}+z_{h_3}$.

One can show that in the leading twist and collinear factorization
approximations,
the semi-inclusive cross section $\sigma_{e^+e^-\rightarrow h_1 h_2 h_3}$
can be expressed in terms of the trihadron fragmentation functions as,
 \begin{equation}
  \frac{d\sigma_{e^+e^-\rightarrow h_1 h_2 h_3}}{dz_{h_1}dz_{h_2}dz_{h_3}}
    =\sum_q\sigma_{0}^{q\bar q}
    \left[D_q^{h_1 h_2 h_3}(z_{h_1},z_{h_2},z_{h_3})
    +D_{\bar q}^{h_1 h_2 h_3}(z_{h_1},z_{h_2},z_{h_3})\right]\, .
 \label{inclusec3}
 \end{equation}

Again, the radiative corrections for the trihadron fragmentation
functions are similar to that of triple quark distribution functions
in Eqs.(\ref{rad31}) and (\ref{rad32}), except that the cutoff
in the intrinsic transverse momentum $\Lambda$ should be replaced
by $Q$ which is the only limit of the relative transverse momenta
between hadrons. The DGLAP equations for the scale evolution for
the trihadron fragmentation functions are then,
 \begin{eqnarray}
   Q^2\frac{\partial}{\partial Q^2} & & \hspace{-0.25in}
   D_q^{h_1 h_2 h_3}(z_{h_1}, z_{h_2}, z_{h_3},Q^2) =
\frac{\alpha_s(Q^2)}{2\pi}
   \left[\int_{z_{h_1}+z_{h_2}+z_{h_3}}^1\frac{dz}{z^3}\gamma_{qq}(z)
   D_q^{h_1 h_2 h_3}\left(\frac{z_{h_1}}{z},\frac{z_{h_2}}{z},
   \frac{z_{h_3}}{z},Q^2\right)\right.
 \nonumber\\
  &+&\sum_{i=1}^3\int_{z_{h_i}}^{1-z_{h_{i^\prime}}-z_{h_{i^{\prime\prime}}}}
   \frac{dz}{z(1-z)^2}{\hat \gamma}_{qq}(z)
   D_q^{h_i}\left(\frac{z_{h_i}}{z},Q^2\right)
   D_g^{h_{i^\prime}h_{i^{\prime\prime}}}\left(\frac{z_{h_{i^\prime}}}{1-z},
   \frac{z_{h_{i^{\prime\prime}}}}{1-z},Q^2\right)
 \nonumber\\
   &+&\sum_{i=1}^3\int_{z_{h_{i^\prime}}+z_{h_{i^{\prime\prime}}}}^{1-z_{h_i}}
   \frac{dz}{z^2(1-z)}{\hat \gamma}_{qq}(z)
   D_q^{h_{i^\prime}h_{i^{\prime\prime}}}\left(\frac{z_{h_{i^\prime}}}{z},
   \frac{z_{h_{i^{\prime\prime}}}}{z},Q^2\right)
   D_g^{h_i}\left(\frac{z_{h_i}}{1-z},Q^2\right)
 \nonumber\\
   &+&\left.\int_{z_{h_1}+z_{h_2}+z_{h_3}}^1\frac{dz}{z^3}\gamma_{qg}(z)
   D_g^{h_1 h_2 h_3}\left(\frac{z_{h_1}}{z},\frac{z_{h_2}}{z},\frac{z_{h_3}}{z},
   Q^2\right)\right]\, ,
 \label{DGLAP31}\\
   Q^2\frac{\partial}{\partial Q^2}& & \hspace{-0.25in}
   D_g^{h_1 h_2 h_3}(z_{h_1}, z_{h_2},  z_{h_3}, Q^2)=
\frac{\alpha_s(Q^2)}{2\pi}
   \left[\int_{z_{h_1}+z_{h_2}+z_{h_3}}^1\frac{dz}{z^3}\gamma_{gq}(z)
   D_s^{h_1 h_2 h_3}\left(\frac{z_{h_1}}{z},\frac{z_{h_2}}{z},
   \frac{z_{h_3}}{z},Q^2\right)\right.
 \nonumber\\
  &+&\sum_{i=1}^3\sum_q\int_{z_{h_i}}^{1-z_{h_{i^\prime}}-z_{h_{i^{\prime\prime}}}}
   \frac{dz}{z(1-z)^2}\gamma_{gq}(z)
   D_q^{h_i}\left(\frac{z_{h_i}}{z},Q^2\right)
   D_{\bar q}^{h_{i^\prime}h_{i^{\prime\prime}}}
   \left(\frac{z_{h_{i^\prime}}}{1-z},
   \frac{z_{h_{i^{\prime\prime}}}}{1-z},Q^2\right)
 \nonumber\\
   &+&\sum_{i=1}^3\sum_q\int_{z_{h_{i^\prime}}
     +z_{h_{i^{\prime\prime}}}}^{1-z_{h_i}}
   \frac{dz}{z^2(1-z)}\gamma_{gq}(z)
   D_q^{h_{i^\prime}h_{i^{\prime\prime}}}\left(\frac{z_{h_{i^\prime}}}{z},
   \frac{z_{h_{i^{\prime\prime}}}}{z},Q^2\right)
   D_{\bar q}^{h_i}\left(\frac{z_{h_i}}{1-z},Q^2\right)
 \nonumber\\
  &+&\int_{z_{h_1}+z_{h_2}+z_{h_3}}^1\frac{dz}{z^3}\gamma_{gg}(z)
   D_g^{h_1 h_2 h_3}\left(\frac{z_{h_1}}{z},\frac{z_{h_2}}{z},\frac{z_{h_3}}{z},
   Q^2\right)
 \nonumber\\
   &+&\sum_{i=1}^3\left.\int_{z_{h_i}}^{1-z_{h_{i^\prime}}-z_{h_{i^{\prime\prime}}}}
   \frac{dz}{z(1-z)^2}{\hat \gamma}_{gg}(z)
   D_g^{h_i}\left(\frac{z_{h_i}}{z},Q^2\right)
   D_g^{h_{i^\prime}h_{i^{\prime\prime}}}\left(\frac{z_{h_{i^\prime}}}{1-z},
   \frac{z_{h_{i^{\prime\prime}}}}{1-z},Q^2\right)\right]\, .
 \label{DGLAP32}
 \end{eqnarray}

\section{Quark Recombination and Jet Fragmentation in a Thermal Medium}
\label{sec5}

So far we have reformulated jet fragmentation in vacuum in the quark 
recombination model in which we related the parton fragmentation 
functions to the
quark recombination probability as determined by the hadron wavefunction
in a constituent quark model and multi-quark distribution functions of
a parton jet before hadronization. In the following sections we will
extend the formula to the case of jet fragmentation in a thermal
medium which is relevant to jet fragmentation in the environment
of high-energy heavy-ion collisions. Such problems have been considered
before \cite{Osborne:2002dx}. But the attention has been focused on parton
emission and absorption by the propagating parton jet in a thermal
medium before hadronization. It has been assumed that the thermal
medium is in a deconfined phase so that partons from the jet will
eventually hadronize together with the medium.

In this paper, however, we focus on the physical process during the
hadronization of the parton jet in a thermal medium and the modification of the
parton fragmentation functions with respective to that in the vacuum.
We assume that the effective degrees of freedom can be described by
constituent quarks just before and during hadronization. Therefore, we 
consider constituent quarks not only as the effective states in the 
process of jet
fragmentation but also the effective constituent of the medium just
before hadronization. In this framework, one naturally encounters
recombination between thermal constituent quarks and shower quarks 
from parton fragmentation, in addition to recombination of shower
quarks as in the vacuum. They both contribute to to hadron production 
associated with an energetic parton jets in a thermal medium.

As in the previous study \cite{Osborne:2002dx}, one can describe 
the fragmentation
of a parton jet in medium simply by replacing the vacuum expectation
in the $S$ matrix of the processes or the operator definition of the
parton fragmentation functions by their thermal expectation values,
$\langle 0|{\cal O} |0\rangle \rightarrow \langle\langle{\cal O}\rangle\rangle$,
 \begin{equation}
 \label{expectation}
  \langle\langle{\cal O}\rangle\rangle=\frac{{\rm Tr}[e^{-{\hat H}\beta}{\cal O}]}
    {{\rm Tr}\;e^{-{\hat H}\beta}}\, ,
 \end{equation}
where, $\hat H$ is the Hamiltonian operator of the system and
$1/\beta=T$ is the temperature. Therefore, the single hadron
fragmentation functions at finite temperature for a quark, antiquark
and gluon jet can be defined as
 \begin{eqnarray}
   {\tilde D_q^h}(z_h, p^+)&=&\frac{z_{h}^3}{2}\int\frac{d^4p}{(2\pi)^4}
   \delta\left(z_{h} {-} \frac{p_{h}^+}{p^+}\right)
   \int d^4 xe^{-ip\cdot x}
    {\rm Tr}\left\lbrack\frac{\gamma^+}{2 p_{h}^+}
    \sum_{S}\langle\langle{\psi}(0)|S,p_{h}\rangle
    \langle p_{h},S|\overline\psi(x)\rangle\rangle\right\rbrack\, ,
 \label{fq11}\\
   {\tilde D}_{\bar q}^h(z_h, p^+)&=&\frac{z_{h}^3}{2}\int\frac{d^4p}{(2\pi)^4}
   \delta\left(z_{h} {-} \frac{p_{h}^+}{p^+}\right)
   \int d^4 xe^{-ip\cdot x}
   {\rm Tr}\left\lbrack
    \sum_{S}\langle \langle{\overline\psi}(0)|S,p_{h}\rangle
    \frac{\gamma^+}{2 p_{h}^+}
    \langle p_{h},S|\psi(x)\rangle\rangle\right\rbrack\, ,
 \label{fq22}\\
   {\tilde D}_g^h(z_h, p^+)&=&\frac{z_{h}^2}{2}\int\frac{d^4p}{(2\pi)^4}
   \delta\left(z_{h}-\frac{p_{h}^+}{p^+}\right)
   \int d^4 xe^{-ip\cdot x}
   d_{\mu\nu}(p)\sum_{S}\langle\langle A^{\mu}(0)
    |S, p_{h}\rangle
    \langle p_{h},S|A^{\nu}(x)\rangle\rangle\, ,
 \label{fg33}
\end{eqnarray}
where $p_h$  and $p$ are the four-momentum of the hadron and the
initial parton, respectively. In the above definition of the parton fragmentation
function in medium, we have explicitly kept the dependence on the
initial parton energy $p^+$. Such dependence arises at finite
temperature from the the thermal average which also introduces
dependence on the temperature $T$. An alternative definition of
the fragmentation functions \cite{Osborne:2002dx},
 \begin{eqnarray}
   {\tilde D_q^h}(p_h^+, p^+)&=&\frac{p^+_h}{4p^+}\int\frac{d^4k}{(2\pi)^4}
   \delta\left(k^+ - p^+\right)
   \int d^4 xe^{-ik\cdot x}
    {\rm Tr}\left\lbrack \gamma^+
    \sum_{S}\langle\langle{\psi}(0)|S,p_{h}\rangle
    \langle p_{h},S|\overline\psi(x)\rangle\rangle\right\rbrack\, ,
 \label{fq1}\\
   {\tilde D}_{\bar q}^h(p_h^+, p^+)&=&\frac{p^+_h}{4p^+}
   \int\frac{d^4k}{(2\pi)^4}
   \delta\left(k^+ - p^+\right)
   \int d^4 xe^{-ik\cdot x}
   {\rm Tr}\left\lbrack
    \sum_{S}\langle \langle{\overline\psi}(0)|S,p_{h}\rangle
    \gamma^+
    \langle p_{h},S|\psi(x)\rangle\rangle\right\rbrack\, ,
 \label{fq2}\\
   {\tilde D}_g^h(p_h^+, p^+)&=&-\frac{p_h^+}{2{p^+}^2}\int\frac{d^4k}{(2\pi)^4}
   \delta\left(k^+ - p^+\right)
   \int d^4 xe^{-ik\cdot x}
  \sum_{S}\langle\langle F^{+\mu}(0)
    |S, p_{h}\rangle
    \langle p_{h},S|F^+_{\mu}(x)\rangle\rangle\, ,
 \label{fg}
\end{eqnarray}
explicitly takes into account the dependence on the
absolute initial parton energy and the medium temperature.
The two expressions can be related via the identity,
\begin{equation}
\frac{p_h^+}{p^+}\delta(k^+ - p^+)=\frac{z_h^3}{p_h^+}\delta(z_h-\frac{p_h^+}{k^+}) .
\end{equation}
At zero temperature $T=0$, the above fragmentation  functions
will be reduced to the parton fragmentation functions
in vacuum as defined in Eqs.(\ref{qfrag1}), (\ref{qfrag2}) and (\ref{gfrag1}).
At leading twist, they depend only on the scaling variable,
$z_h=p_h^+/p^+$, {\it i.e.}, the ratio of the hadron and parton energies.

\subsection{Meson Production from Thermal Quark Recombination}
\label{sec5a}

From Eq. (\ref{fq11}), the meson fragmentation function at finite temperature
from the quark parton can be expressed as
 \begin{eqnarray}
   {\tilde D}_q^M(z_M, p^+)&=&\frac{z_{M}^3}{2}\int\frac{d^4p}{(2\pi)^4}
   \delta\left(z_{M} {-} \frac{p_{M}^+}{p^+}\right)
   \int d^4 xe^{-ip\cdot x}
 \nonumber\\
   &&\int[d^2k_{\perp}][dx]
   \int[d^2k'_{\perp}][dx']
   \varphi_M(k_{1\perp}, x_1; k_{2\perp}, x_2)
   \varphi^*_M(k'_{1\perp}, x'_1; k'_{2\perp}, x'_2)
  \nonumber\\
   &&{\rm Tr}\left[\frac{\gamma^+}{2 p_{M}^+}\sum_{ \widetilde S}
   \langle\langle\psi(0)|\widetilde S; k_{1\perp}, x_1; k_{2\perp}, x_2\rangle
   \langle k'_{2\perp}, x'_2; k'_{1\perp}, x'_1; \widetilde S
   |\overline\psi(x)\rangle\rangle\right] \, .
 \label{fm}
\end{eqnarray}

To evaluate the thermal average in Eq.~(\ref{fm}), it is
convenient to use the finite volume quantization (FVQ) in a cubic box
with length $L$. Employing FVQ, the momentum integral, $\delta$-function 
and  continuous states can be replaced by following summation, Kronecker 
functions and discrete states, respectively.
 \begin{eqnarray}
   \int_{-\infty}^{\infty}dq_x\int_{-\infty}^{\infty}dq_y
   \int_{-\infty}^{\infty}dq_z f(q_x,q_y,q_z)
   &\longleftrightarrow&
   \left(\frac{2\pi}{L}\right)^3\sum_{n_x=-\infty}^{\infty}
   \sum_{n_y=-\infty}^{\infty}\sum_{n_z=-\infty}^{\infty}
   f(\frac{2\pi n_x}{L},\frac{2\pi n_y}{L},\frac{2\pi n_z}{L})\, ,
 \label{replace1}\\
   (2\pi)^3\delta(q_x-q'_x)\delta(q_y-q'_y)\delta(q_z-q'_z)
   &\longleftrightarrow&
   L^3\delta_{n_x,n'_x}\delta_{n_y,n'_y}\delta_{n_z,n'_z}\, ,
 \label{replace2}\\
   \frac{1}{\sqrt{2E_q L^3}}|q\rangle
   &\longleftrightarrow& |q_i \rangle\, .
 \label{replace3}
 \end{eqnarray}

Using Eq.(\ref{expectation}), we can express the thermal average
of the matrix element in Eq.~(\ref{fm}) as
 \begin{eqnarray}
   &&\langle\langle\psi(0)|\widetilde S; k_{1\perp}, x_1; k_{2\perp}, x_2\rangle
   \langle  k'_{2\perp}, x'_2; k'_{1\perp}, x'_1; \widetilde S
   |\overline\psi(x)\rangle\rangle
 \nonumber\\
   &&\qquad=\frac{
   \underset{\{n_i\}}{\sum}\langle
   \{(q_i,n_i)\}|e^{-\beta \hat H}\psi(0)|\widetilde S; k_1; k_2\rangle
   \langle k'_2; k'_1; \widetilde S|\overline\psi(x)|\{(q_i,n_i)\}\rangle}
   {\underset{\{n_i\}}{\sum}\langle
   \{(q_i,n_i)\}|e^{-\beta \hat H}|\{(q_i,n_i)\}\rangle}\, ,
 \label{average1}
 \end{eqnarray}
where $n_i=0, 1$ (for $i=1, 2\cdots,\infty$) since we have assumed the
thermal medium to consist of constituent quarks that obey Fermi statistics.
The denominator in Eq.(\ref{average1}) can be reduced to
 \begin{eqnarray}
   &&\sum_{\{n_i\}}\langle
   \{(q_i,n_i)\}|e^{-\beta \hat H}|\{(q_i,n_i)\}\rangle
   =\sum_{\{n_i\}}e^{-\beta\overset{\infty}{\underset{i=1}{\sum}}n_i E_i}
   =\prod_{i=1}^{\infty}\left( 1+e^{-\beta E_i}\right)\, .
 \label{denom}
 \end{eqnarray}
Taking into account that the intermediate states of constituent quarks
$|k_1\rangle, |k_2\rangle, |k'_1\rangle, |k'_2\rangle$ can
contract both with the field operators $\psi$ and $\overline\psi$ and the
constituent quark states $|\{(q_i,n_i)\}\rangle$ in the thermal medium,
we get
 \newcommand{\RR}{\rule{.4pt}{3pt}}
 \begin{eqnarray}
   &&\langle\langle\psi(0)|\widetilde S; k_{1\perp}, x_1; k_{2\perp}, x_2\rangle
   \langle  k'_{2\perp}, x'_2; k'_{1\perp}, x'_1; \widetilde S
   |\overline\psi(x)\rangle\rangle
 \nonumber\\
    &&\qquad=\left(\prod_{i=1}^{\infty}\left( 1+e^{-\beta E_i}\right)\right)^{-1}
    \sum_{\{n_i\}}e^{-\beta\overset{\infty}{\underset{i=1}{\sum}}n_i E_i}
  \nonumber\\
    &&\qquad\quad \Bigl[\langle
    (q_1,n_1);\cdots ;(q_{\infty},n_{\infty})|
    \psi(0)|\widetilde S; k_1; k_2\rangle\langle k'_2; k'_1;
    \widetilde S|\overline\psi(x)|(q_{\infty},n_{\infty});\cdots;
    (q_1,n_1)\rangle
  \nonumber\\
     &&\qquad\quad +{\underset{i,j}{\sum}}\langle (q_1,n_1);
    \cdots ;(\underset{\RR\hrulefill\RR}{q_i,n_i); \cdots|
    \psi(0)|\widetilde S; k_1}; k_2\rangle
    \langle k'_2;\underset{\RR\hrulefill\RR}{k'_1;
    \widetilde S|\overline\psi(x)|\cdots;
    (q_j},n_j);\cdots; (q_1,n_1)\rangle
  \nonumber\\
     &&\qquad\quad +{\underset{i,j}{\sum}}\langle (q_1,n_1);
    \cdots ;(\underset{\RR\hrulefill\RR}{q_i,n_i); \cdots|
    \psi(0)|\widetilde S; k_1; k_2}\rangle
    \langle \underset{\RR\hrulefill\RR}{k'_2;k'_1;
    \widetilde S|\overline\psi(x)|\cdots;
    (q_j},n_j);\cdots; (q_1,n_1)\rangle
  \nonumber\\
    &&\qquad\quad +{\underset{i,j,l,m}{\sum}}\langle \cdots; (\underset{\RR\hrulefill\RR}{q_i,n_i);
    \cdots ;(\underset{\RR\hrulefill\RR}{q_l,n_l); \cdots|
    \psi(0)|\widetilde S; k_1}; k_2}\rangle
    \langle \underset{\RR\hrulefill\RR}{k'_2;\underset{\RR\hrulefill\RR}
    {k'_1; \widetilde S|\overline\psi(x)|\cdots;
    (q_m},n_m);\cdots; (q_j},n_j); \cdots\rangle\Bigr]\, .
 \label{average2}
 \end{eqnarray}
For simplicity in this paper, the states $|k_1\rangle,
|k_2\rangle, |k'_1\rangle, |k'_2\rangle$ without the connecting
lines are meant to be only contracted with the field operators
$\psi$ and $\overline\psi$. The states $|(q_i,n_i)\rangle$ with
the connecting lines are meant to contract only with the final
states $|k_{1,2}\rangle$ and $|k'_{1,2}\rangle$ but without
contraction with the field operators $\psi$ and $\overline\psi$.
The states $|(q_i,n_i)\rangle$ without the connecting lines can
still contract with the field operators $\psi$ and $\overline\psi$
but will not contract with the final states $|k_{1,2}\rangle$ and
$|k'_{1,2}\rangle$.

Let us for a moment neglect the contraction between the thermal
states $|(q_i,n_i)\rangle$ and the field operators $\psi$ and
$\overline\psi$. Using
 \begin{eqnarray}
    \langle (q_1,n_1); \cdots ;(q_{i-1},n_{i-1}); (q_{i+1},n_{i+1});\cdots|
    \cdots; (q_{j+1},n_{j+1});(q_{j-1},n_{j-1});
    \cdots; (q_1,n_1)\rangle=\delta_{ij}\, ,
 \label{normalization}
 \end{eqnarray}
one can obtain from Eq.~(\ref{average2})
 \begin{eqnarray}
   &&\langle\langle\psi(0)|\widetilde S; k_{1\perp}, x_1; k_{2\perp}, x_2\rangle
   \langle  k'_{2\perp}, x'_2; k'_{1\perp}, x'_1; \widetilde S
   |\overline\psi(x)\rangle\rangle
 \nonumber\\
   &&\qquad=\left(\prod_{i=1}^{\infty}\left( 1+e^{-\beta E_i}\right)\right)^{-1}\Bigl[
   \sum_{\{n_i\}}
   e^{-\beta\overset{\infty}{\underset{i=1}{\sum}}n_i E_i}\langle 0|
   \psi(0)|\widetilde S; k_1; k_2\rangle
   \langle k'_2; k'_1; \widetilde S|\overline\psi(x)|0\rangle
 \nonumber\\
   &&\qquad +{\underset{j}{\sum}}
   e^{-\beta E_j}\sideset{}{'}\sum_{\{n_i\}}
   e^{-\beta\overset{\infty}{\underset{i=1}{\sideset{}{'}\sum}}n_i E_i}
   \langle (\underset{\RR\hrulefill\RR}{q_j,n_j)|
   \psi(0)|\widetilde S; k_1}; k_2\rangle
   \langle k'_2;\underset{\RR\hrulefill\RR}
   {k'_1; \widetilde S|\overline\psi(x)|(q_j},n_j)\rangle
 \nonumber\\
   &&\qquad +{\underset{j}{\sum}}
   e^{-\beta E_j}\sideset{}{'}\sum_{\{n_i\}}
   e^{-\beta\overset{\infty}{\underset{i=1}{\sideset{}{'}\sum}}n_i E_i}
   \langle (\underset{\RR\hrulefill\RR}{q_j,n_j)|
   \psi(0)|\widetilde S; k_1; k_2}\rangle
   \langle \underset{\RR\hrulefill\RR}
   {k'_2;k'_1; \widetilde S|\overline\psi(x)|(q_j},n_j)\rangle
 \nonumber\\
   &&\qquad +{\underset{j,m}{\sum}}
   e^{-\beta (E_j+E_m)}\sideset{}{''}\sum_{\{n_i\}}
   e^{-\beta\overset{\infty}{\underset{i=1}{\sideset{}{''}\sum}}n_i E_i}
   \langle (\underset{\RR\hrulefill\RR}{q_j,n_j);(\underset{\RR\hrulefill\RR}{q_m,n_m)|
   \psi(0)|\widetilde S; k_1}; k_2}\rangle
   \langle \underset{\RR\hrulefill\RR}{k'_2;\underset{\RR\hrulefill\RR}
   {k'_1; \widetilde S|\overline\psi(x)|(q_m},n_m);(q_j},n_j)\rangle\Bigr]\, ,
 \label{average3}
 \end{eqnarray}
where $\Sigma^{'}, \Sigma^{''} $ indicates that $i\neq j$ and
$i\neq j, m$ in their summation, respectively. One can generalize
the above results to include the contraction between thermal
states and the quark field operators. In this general case, the
vacuum expectation will be replaced by a thermal average
without contraction between thermal states and the final
constituent quark states. Using Eq.(\ref{denom}) and replacements
(\ref{replace1}),(\ref{replace3}), we obtain the thermal average
value of the matrix elements in the meson fragmentation function
as
 \begin{eqnarray}
   &&\langle\langle\psi(0)|\widetilde S; k_{1\perp}, x_1; k_{2\perp}, x_2\rangle
   \langle  k'_{2\perp}, x'_2; k'_{1\perp}, x'_1; \widetilde S
   |\overline\psi(x)\rangle\rangle
   =\,_k\langle\langle
   \psi(0)|\widetilde S; k_1; k_2\rangle
  \langle k'_2;k'_1; \widetilde S|\overline\psi(x)\rangle\rangle_k
 \nonumber\\
   &&\quad +
   \int\frac{d^3q_1}{(2\pi)^3 2E_{q_1}}
   \frac{1}{e^{\beta E_{q_1}}+1}\Bigl[
   \,_k\langle\langle \underset{\RR\hrulefill\RR}{q_1|
   \psi(0)|S; k_1}; k_2\rangle
   \langle k'_2;\underset{\RR\hrulefill\RR}{k'_1; \widetilde S|\overline\psi(x)|q_1}
   \rangle\rangle_k
   +
   \,_k\langle\langle \underset{\RR\hrulefill\RR}{q_1|
   \psi(0)|\widetilde S; k_1; k_2}\rangle
   \langle\underset{\RR\hrulefill\RR}{k'_2;k'_1; \widetilde S|\overline\psi(x)|q_1}
   \rangle\rangle_k\Bigr]
 \nonumber\\
   &&\quad +
   \int\frac{d^3q_1}{(2\pi)^3 2E_{q_1}}\frac{d^3q_2}{(2\pi)^3 2E_{q_2}}
   \frac{1}{e^{\beta E_{q_1}}+1}\frac{1}{e^{\beta E_{q_2}}+1}
   \,_k\langle\langle\underset{\RR\hrulefill\RR}{q_2 ;\underset{\RR\hrulefill\RR}{q_1|
   \psi(0)|\widetilde S; k_1}; k_2}\rangle
   \langle \underset{\RR\hrulefill\RR}{k'_2;\underset{\RR\hrulefill\RR}
   {k'_1; \widetilde S|\overline\psi(x)|q_1};q_2}\rangle\rangle_k\, ,
 \label{average4}
 \end{eqnarray}
where the notation $\,_k\langle\langle \cdots \rangle\rangle_k$ represents
thermal averaging without the contraction between thermal states and the
final constituent quark states. The form $(e^{\beta E}+1)^{-1}$ is the 
Fermi-Dirac thermal distribution function
in the co-moving frame of a fluid that maintains only local equilibrium. In
the reaction frame in which a parton jet propagates, it should be replaced by
 \begin{eqnarray}
  f(q)&=&\frac{1}{e^{\beta q\cdot u}+1}\, ,
 \label{distribution1}
 \end{eqnarray}
for the thermal quark distribution in a fluid element with flow 
four-velocity $u$
in the thermal medium. The momentum of a thermal constituent quark
is decomposed to $q=(q^+,0,q_\perp)$ with respect to the direction of the
propagating parton jet. In the following, we also sometimes express the
longitudinal component of the thermal momentum as a fraction of the initial
parton momentum, $q^+=z_qp^+$. In this case, the value of $z_q$ is not bound.
With this, we can finally express the thermal averaged matrix element in the
fragmentation function as
 \begin{eqnarray}
   &&\langle\langle\psi(0)|\widetilde S; k_{1\perp}, x_1; k_{2\perp}, x_2\rangle
   \langle  k'_{2\perp}, x'_2; k'_{1\perp}, x'_1; \widetilde S
   |\overline\psi(x)\rangle\rangle
   =\,_k\langle\langle |\psi(0)|\widetilde S; k_1; k_2\rangle
   \langle k'_2;k'_1; \widetilde S|\overline\psi(x)|\rangle\rangle_k
 \nonumber\\
   &&\quad +
   \int\frac{d^2q_{1\perp}}{2(2\pi)^3}\frac{dq_1^+}{q_1^+}
   f(q_{1\perp}, q_1^+)\Bigl[
   \,_k\langle\langle \underset{\RR\hrulefill\RR}{q_1|
   \psi(0)|S; k_1}; k_2\rangle
   \langle k'_2;\underset{\RR\hrulefill\RR}
   {k'_1; \widetilde S|\overline\psi(x)|q_1}\rangle\rangle_k
    +\,_k\langle\langle \underset{\RR\hrulefill\RR}{q_1|
   \psi(0)|\widetilde S; k_1; k_2}\rangle
   \langle \underset{\RR\hrulefill\RR}{k'_2;k'_1; \widetilde S|\overline\psi(x)|q_1}
   \rangle\rangle_k\Bigr]
 \nonumber\\
   &&\quad +
   \int\frac{d^2q_{1\perp}}{2(2\pi)^3}\frac{dq_1^+}{q_1^+}
   \int\frac{d^2q_{2\perp}}{2(2\pi)^3}\frac{dq_2^+}{q_2^+}
   f(q_{1\perp}, q_1^+)f(q_{2\perp}, q_2^+)
   \,_k\langle\langle\underset{\RR\hrulefill\RR}{q_2 ;\underset{\RR\hrulefill\RR}{q_1|
   \psi(0)|\widetilde S; k_1}; k_2}\rangle
   \langle \underset{\RR\hrulefill\RR}{k'_2;\underset{\RR\hrulefill\RR}
   {k'_1; \widetilde S|\overline\psi(x)|q_1};q_2}\rangle\rangle_k\, .
 \label{average}
 \end{eqnarray}

 \begin{figure}
\resizebox{2.5in}{1.5in}{\includegraphics{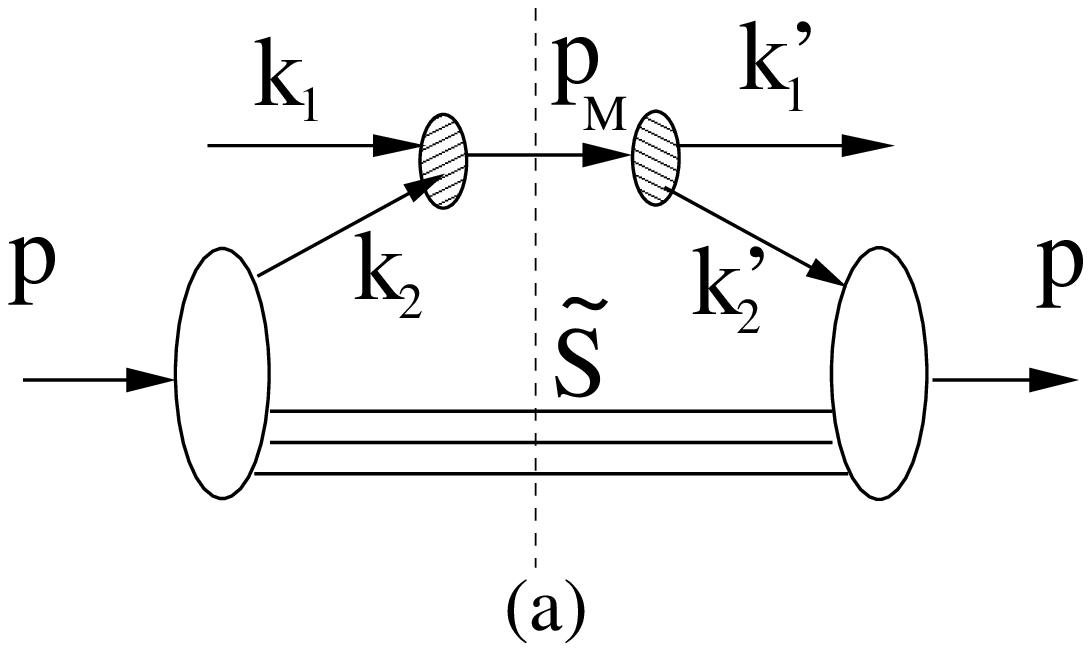}}
\hspace{0.5in}
\resizebox{2.5in}{1.5in}{\includegraphics{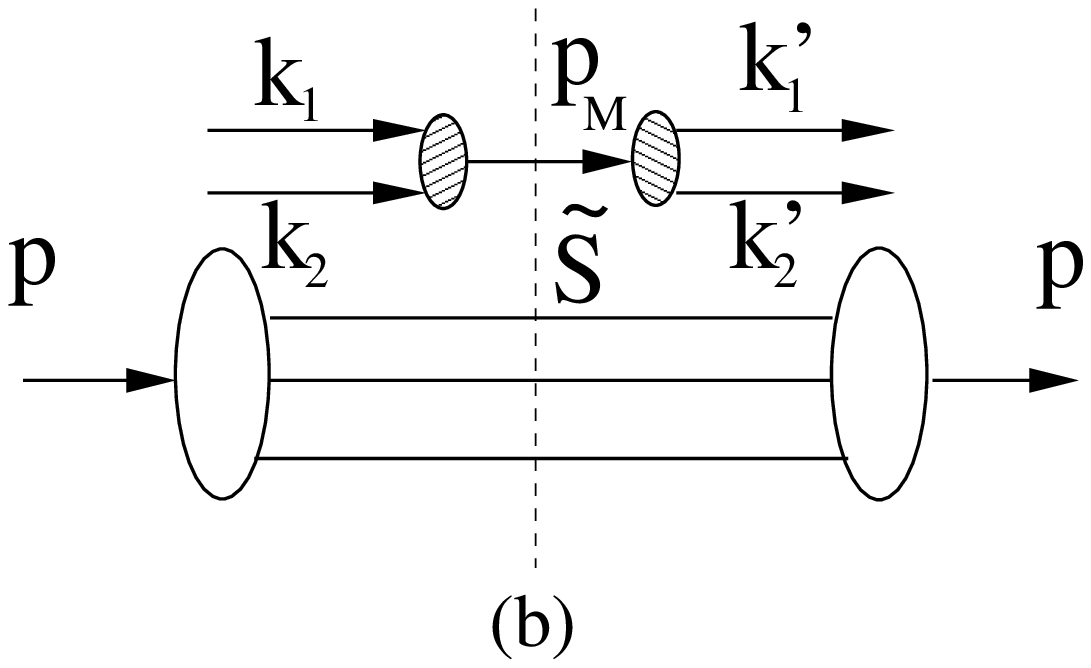}}
 \caption{The cut diagram for contributions from (a) shower-thermal
and (b) thermal-thermal quark recombination to the single meson fragmentation
function.}
 \label{fig4}
 \end{figure}

The above equation has three distinct contributions. One can immediately
identify the first term as the contribution from recombination of two
shower quarks from parton fragmentation,
 \begin{eqnarray}
  {\tilde D}_{q}^{M(SS)}(z_M)
  \approx C_M \int_0^{z_M}\frac{dz_1}{2}
  R_M(0_\perp,\frac{z_1}{z_M})
  {\tilde F}_{q}^{q_1{\bar q}_2}(z_1, z_M-z_1)  \, ,
 \label{qfragm03}
 \end{eqnarray}
This contribution is normally referred \cite{Hwa:2002tu} to as
``shower-shower" quark recombination. It has exactly the same form
as the parton fragmentation functions in vacuum [Eq.(\ref{qfragm02})]
in the framework of quark recombination, except that the diquark
distribution functions ${\tilde F}_{q}^{q_1{\bar q}_2}(z_1, z_2)$ are
now medium modified. Their definitions are similar to that in vacuum
in Eq.~(\ref{qfrag03}) but the vacuum expectation is replaced by a thermal
average. These modified diquark distribution functions should in principle
contain effects of multiple scattering, induced gluon radiation and
absorption, in the same way as the modification of hadron fragmentation
functions in a thermal medium \cite{Osborne:2002dx}. We have assumed that
the meson wavefunction and therefore the recombination probability
during the hadronization of the parton jets in a quark-gluon plasma are
the same as in vacuum.

In the second term, one of the
constituent quarks is contracted with the thermal quark states while the
other with the quark field operators. This corresponds to processes in
which a quark from the thermal medium combines with another quark from
the parton fragmentation (or shower) to form the final meson, as
illustrated in Fig.~\ref{fig4}(a). Using the convention of previous 
studies \cite{Hwa:2002tu}, these processes are called
``shower-thermal" quark recombination. Finally, the third term
where both of the final constituent quark states contract with the
thermal quark states corresponds to the formation of the final
meson from two thermal quarks in the quark recombination model,
as shown in Fig.~\ref{fig4}(b). This is referred to as  ``thermal-thermal"
quark recombination.

Using Eqs.~(\ref{Rm}) and (\ref{fm}) and the contraction between
two single-particle states,
 \begin{eqnarray}
   \langle\underset{\RR\hrulefill\RR}{q_i|k_j}\rangle \equiv
   \langle q_{i\perp}, x_{qi}|k_{j\perp}, x_j\rangle
   =(2\pi)^3 2x_i \delta^{(2)}\left(q_{i\perp}-k_{j\perp}\right)
  \delta\left(x_{qi}-x_j\right)\, ,
  \label{contract}
 \end{eqnarray}
we can express the contribution from ``shower-thermal'' quark recombination
to the meson fragmentation function as
 \begin{eqnarray}
  {\tilde D}_{q}^{M(ST)}(z_M, p^+)
  &=&\int_0^{z_M} \frac{dz_q}{z_M} \int\frac{d^2q_{\perp}}{2(2\pi)^3}
  R_M(q_\perp, \frac{z_q}{z_M})
  \frac{1}{(1-z_q/z_M)^2}
  \nonumber \\
  & & \left [ f_{q_1}(q_{\perp},z_q p^+){\tilde F}_{q}^{\bar q_2}(z_M-z_q)
  +f_{\bar q_2}(q_{\perp},z_q p^+){\tilde F}_{q}^{q_1}(z_M-z_q) \right ]
  \, ,
 \label{mesonST}
 \end{eqnarray}
where $x_{q_i}=z_{q_i}/z_M$ is the thermal quark's momentum as a fraction 
of the produced meson's momentum,
${\tilde F}_{q}^{q_i}(z)$ and ${\tilde F}_{q}^{\bar q_i}(z)$ are single
constituent quark or anti-quark distributions of the fragmenting parton jet
in a thermal medium
defined similarly as in Eqs.~(\ref{qdistr1}) and (\ref{qbardistr}),
except that the vacuum expectation values are replaced again by thermal
averaged expectation. They should be different from the
corresponding quark distributions in vacuum because of multiple scattering,
induced gluon bremsstrahlung and parton absorption.

The contribution due to ``thermal-thermal'' quark recombination to the
meson fragmentation function in Eq.~(\ref{average}) comes from
recombination of two thermal constituent quarks. This contribution
would never be associated with the parton jet had we not used the
parton's momentum as a reference to calculate the distribution of
produced hadrons from such thermal quark recombination.
Again, using Eq.~(\ref{contract}), we can express this ``thermal-thermal''
contribution as
 \begin{eqnarray}
   {\tilde D}_{q}^{M(TT)}(z_M, p^+)
   &=&\int\frac{d^2q_{1\perp}dz_{q_1}}{2(2\pi)^3}
   \int\frac{d^2q_{2\perp}dz_{q_2}}{2(2\pi)^3}
   f_{q_1}(q_{1\perp}, z_{q_1}p^+)f_{\bar q_2}(q_{2\perp}, z_{q_2}p^+)
   S_{q}^0(z_M)
 \nonumber\\
   &\times& 4 \left|\varphi_M (q_{1\perp}, \frac{z_{q1}}{z_M}; q_{2\perp},
   \frac{z_{q2}}{z_M})\right|^2
   \left[(2\pi)^3\delta^{(2)}(p_{M\perp}-q_{1\perp}-q_{2\perp})
   \delta(z_M-z_{q_1}-z_{q_2})\right]^2 \, ,
 \label{mesonTT1}
 \end{eqnarray}
where
 \begin{eqnarray}
  S_{q}^{0}(z_{M})
   &\equiv&\frac{z_{M}^3}{2}\int\frac{d^4p}{(2\pi)^4}
   \delta\left(z_{M} {-} \frac{p_{M}^+}{p^+}\right)
   \int d^4 xe^{-ip\cdot x}
   {\rm Tr}\left\lbrack\frac{\gamma^+}{2 p_{M}^+}
   \sum_{\widetilde S}\,_k\langle\langle \left|{\psi}(0)
   \right|\widetilde S\rangle \langle \widetilde S\left|\overline\psi(x)
   \right|0\rangle\rangle_k\right\rbrack
 \nonumber\\
   &=&\frac{z_M}{2}\frac{1}{(2\pi)^3}\int d^2p_{\perp}
   = \frac{1}{2(2\pi)^3}\frac{1}{z_M}\int d^2p_{M\perp}\, ,
 \label{dq0}
 \end{eqnarray}
is just a factor associated with the phase-space integration.
Since the fragmentation function represents the total yield of
particle production, the thermal contribution naturally contains a
volume factor, which leads to the $\delta$-function squared in
Eq.~(\ref{mesonTT1}). To factor out this volume
factor, we again employ the finite volume quantization with $V=L^3$ in
a cubic box. The momentum integral and $\delta$-functions can be
expressed by using replacements Eqs. ~(\ref{replace1}) and
(\ref{replace2}). Then we can rewrite Eq.~(\ref{mesonTT1}) as
 \begin{eqnarray}
   {\tilde D}_{q}^{M(TT)}(z_M, p^+)
   &=& Vp^+ \int\frac{d^2p_{M\perp}}{(2\pi)^3}
   \int_0^{z_M}\frac{dz_q}{2z_M}\int\frac{d^2q_{\perp}}{(2\pi)^3}
 \nonumber\\
   &&f_{q_1}(q_{\perp}, z_{q}p^+)f_{\bar q_2}(p_{M\perp}-q_{\perp}, (z_M-z_q)p^+)
   R_M(q_{\perp}, \frac{z_q}{z_M})\, .
 \label{mesonTT2}
 \end{eqnarray}

Since hadron production from the thermal quark recombination is not
correlated with the parton jet and its fragmentation, the above expression
is a little unnatural. One should be able to rewrite it in a form
that has no dependence on the parton jet. Considering,
 \begin{eqnarray}
   {\tilde D}_{q}^{M(TT)}(z_M, p^+)=\frac{dN^{M(TT)}}{dz_M}
   =\frac{dN^{M(TT)}}{dp_M^+} p^+\, ,
 \label{mesonTT3}
 \end{eqnarray}
one can obtain the invariant hadron spectrum from thermal quark
recombination,
 \begin{equation}
   (2\pi)^3\frac{dN^{M(TT)}}{dp_M^+d^2p_{M\perp}}
   =V\int_0^1 dx_q\int\frac{d^2q_{\perp}}{2(2\pi)^3}
   f_{q_1}(q_{\perp}, x_{q}p_M^+)f_{\bar q_2}(p_{M\perp}-q_{\perp}, (1-x_q)p_M^+)
R_M(q_{\perp}, x_{q})\, .
 \label{mesonTT4}
 \end{equation}
which coincides with results from other recombination
models \cite{Greco:2003xt,Fries:2003vb}.
In this expression, the hadron spectra from  thermal quark recombination
are not correlated and therefore do not depend on the parton jet
fragmentation. Even though one can sum over the three contributions and
obtain the effective meson fragmentation in a thermal medium,
 \begin{eqnarray}
   {\tilde D}_{q}^{M}(z_M, p^+)={\tilde D}_q^{M(SS)}(z_M)
   +{\tilde D}_{q}^{M(ST)}(z_M, p^+)+{\tilde D}_{q}^{M(TT)}(z_M, p^+)\, ,
 \label{mesonF}
 \end{eqnarray}
the last term from thermal quark recombination is not correlated with
the initial parton jet and therefore should not be considered as
part of the medium modified jet fragmentation function. The
contributions that are correlated with the initial parton jets
are from ``shower-shower'' and ``shower-thermal'' quark recombination.
For a thermalized medium, the thermal quark distribution follows a
Fermi-Dirac form that is determined by the local temperature and flow
velocity. The contribution from ``shower-thermal'' recombination
is most important for hadron spectra in the intermediate transverse
momentum region. This contribution, however, will be negligible
relative to the ``shower-shower'' recombination which dominates hadron
spectra at large transverse momentum because that the power-law-like
spectra of initially produced partons will win over the
exponential-like distribution of thermal quarks.

\subsection{Baryon Production from Thermal Quark Recombination}
\label{sec5b}

We can similarly generalize the quark recombination model for
baryon fragmentation functions to the case in a thermal medium,
 \begin{eqnarray}
   {\tilde D}_q^B(z_B, p^+)&=&\frac{z_{B}^3}{2}\int\frac{d^4p}{(2\pi)^4}
   \delta\left(z_{B} {-} \frac{p_{B}^+}{p^+}\right)
   \int d^4 xe^{-ip\cdot x}
 \nonumber\\
   &&\int\frac{[d^2k_{\perp}][dx]}{\sqrt{x_1}\sqrt{x_2}\sqrt{x_3}}
   \int\frac{[d^2k'_{\perp}][dx']}{\sqrt{x'_1}\sqrt{x'_2}\sqrt{x'_3}}
   \varphi_B(k_{1\perp}, x_1; k_{2\perp}, x_2; k_{3\perp}, x_3)
   \varphi^*_B(k'_{1\perp}, x'_1; k'_{2\perp}, x'_2; k'_{3\perp}, x'_3)
  \nonumber\\
   &&{\rm Tr}\left[\frac{\gamma^+}{2 p_{B}^+}\sum_{\widetilde S}
   \langle\langle\psi(0)|\widetilde S; k_{1\perp}, x_1;
   k_{2\perp}, x_2; k_{3\perp}, x_3\rangle
   \langle k'_{3\perp}, x'_3; k'_{2\perp}, x'_2; k'_{1\perp}, x'_1; 
   \widetilde S
   |\overline\psi(x)\rangle\rangle\right] \, .
 \label{fb}
\end{eqnarray}
As in the case of meson fragmentation functions, the thermal average of
the matrix element in the above baryon fragmentation function can be
expressed as
 \begin{eqnarray}
   &&\langle\langle\psi(0)|\widetilde S; k_{1\perp}, x_1; k_{2\perp}, x_2; k_{3\perp}, x_3\rangle
   \langle k'_{3\perp}, x'_3; k'_{2\perp}, x'_2; k'_{1\perp}, x'_1; \widetilde S
   |\overline\psi(x)\rangle\rangle
 \nonumber\\
  &&\quad =\, _k\langle\langle\psi(0)|\widetilde S; k_1; k_2;k_3\rangle
   \langle k'_3;k'_2;k'_1;\widetilde S|\overline\psi(x)\rangle\rangle_k
 \nonumber\\
   &&\quad +
   \int\frac{d^2q_{1\perp}}{2(2\pi)^3}\frac{dq_1^+}{q_1^+}
   f(q_{1\perp}, q_1^+)\Bigl[
   \, _k\langle\langle \underset{\RR\hrulefill\RR}{q_1|
   \psi(0)|\widetilde S; k_1}; k_2; k_3\rangle
   \langle k'_3; k'_2;\underset{\RR\hrulefill\RR}
   {k'_1; \widetilde S|\overline\psi(x)|q_1}\rangle\rangle_k
 \nonumber\\
   &&\quad +\, _k\langle\langle \underset{\RR\hrulefill\RR}{q_1|
   \psi(0)|\widetilde S; k_1; k_2};k_3\rangle
 \langle k'_3; \underset{\RR\hrulefill\RR}{k'_2;k'_1; \widetilde S|
     \overline\psi(x)|q_1} \rangle\rangle_k
   +\, _k\langle\langle \underset{\RR\hrulefill\RR}{q_1|
   \psi(0)|\widetilde S; k_1; k_2;k_3}\rangle
   \langle \underset{\RR\hrulefill\RR}{k'_3; k'_2;k'_1; \widetilde S|\overline\psi(x)|q_1}
   \rangle\rangle_k\Bigr]
 \nonumber\\
   &&\quad +
   \int\frac{d^2q_{1\perp}}{2(2\pi)^3}\frac{dq_1^+}{q_1^+}
   \int\frac{d^2q_{2\perp}}{2(2\pi)^3}\frac{dq_2^+}{q_2^+}
   f(q_{1\perp}, q_1^+)f(q_{2\perp}, q_2^+)
   \Bigl[\, _k\langle\langle\underset{\RR\hrulefill\RR}{q_2 ;\underset{\RR\hrulefill\RR}{q_1|
   \psi(0)|\widetilde S; k_1}; k_2};k_3\rangle
   \langle k'_3; \underset{\RR\hrulefill\RR}{k'_2;\underset{\RR\hrulefill\RR}
   {k'_1;\widetilde S|\overline\psi(x)|q_1};q_2}\rangle\rangle_k
 \nonumber\\
   &&\quad+\, _k\langle\langle\underset{\RR\hrulefill\RR}{q_2 ;\underset{\RR\hrulefill\RR}{q_1|
   \psi(0)|\widetilde S; k_1}; k_2;k_3}\rangle
   \langle \underset{\RR\hrulefill\RR}{k'_3; k'_2;\underset{\RR\hrulefill\RR}
   {k'_1;\widetilde S|\overline\psi(x)|q_1};q_2}\rangle\rangle_k
   +\, _k\langle\langle\underset{\RR\hrulefill\RR}{q_2 ;\underset{\RR\hrulefill\RR}{q_1|
   \psi(0)|\widetilde S; k_1; k_2};k_3}\rangle
   \langle \underset{\RR\hrulefill\RR}{k'_3; \underset{\RR\hrulefill\RR}
   {k'_2;k'_1;\widetilde S|\overline\psi(x)|q_1};q_2}\rangle\rangle_k\Bigr]
  \nonumber\\
   &&\quad +
   \int\frac{d^2q_{1\perp}}{2(2\pi)^3}\frac{dq_1^+}{q_1^+}
   \int\frac{d^2q_{2\perp}}{2(2\pi)^3}\frac{dq_2^+}{q_2^+}
   \int\frac{d^2q_{3\perp}}{2(2\pi)^3}\frac{dq_3^+}{q_3^+}
 \nonumber\\
   &&\qquad f(q_{1\perp}, q_1^+)f(q_{2\perp}, q_2^+)f(q_{3\perp}, q_3^+)
   \, _k\langle\langle\underset{\RR\hrulefill\RR}{q_3 ;
   \underset{\RR\hrulefill\RR}{q_2 ;\underset{\RR\hrulefill\RR}{q_1|
   \psi(0)|\widetilde S; k_1}; k_2};k_3}\rangle
   \langle \underset{\RR\hrulefill\RR}{k'_3;
   \underset{\RR\hrulefill\RR}{k'_2;\underset{\RR\hrulefill\RR}
   {k'_1;\widetilde S|\overline\psi(x)|q_1};q_2};q_3}\rangle\rangle_k\, .
 \label{averageb}
 \end{eqnarray}
Here, one can similarly identify the first term
and the last term as
the contributions from "shower-shower-shower" and "thermal-thermal-thermal" quark
recombination, respectively. The others terms are the contributions from
the "shower--shower-thermal" and ``shower-thermal-thermal'' quark
recombination. These last three processes involving thermal quarks
are illustrated in Fig.~\ref{fig5}.

 \begin{figure}
\resizebox{2.5in}{1.5in}{\includegraphics{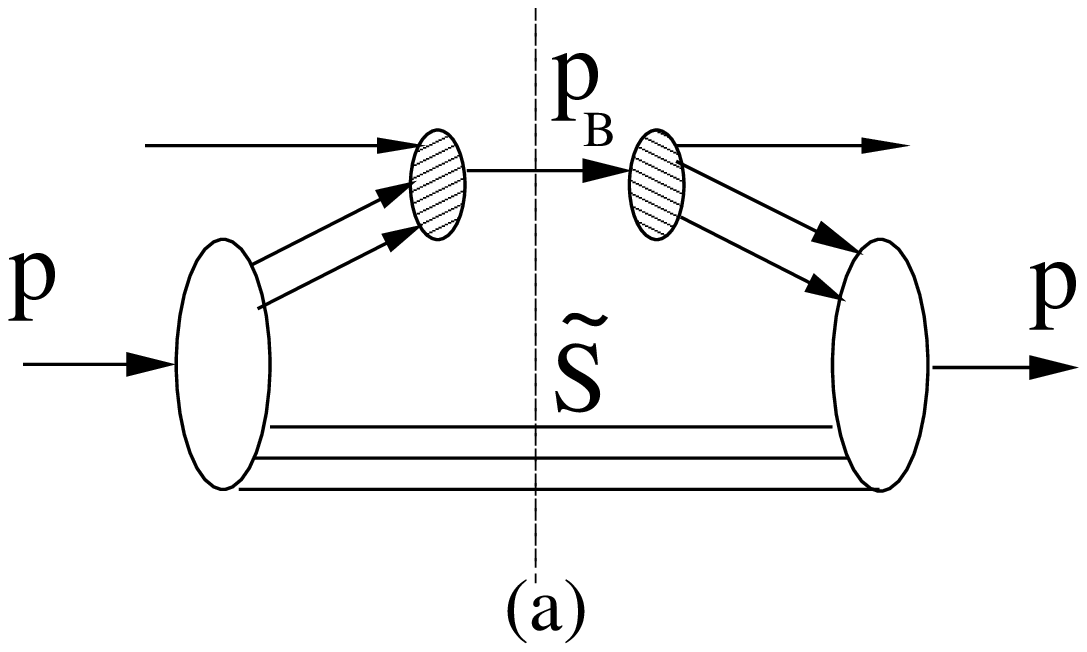}}
\hspace{0.5in}
\resizebox{2.5in}{1.5in}{\includegraphics{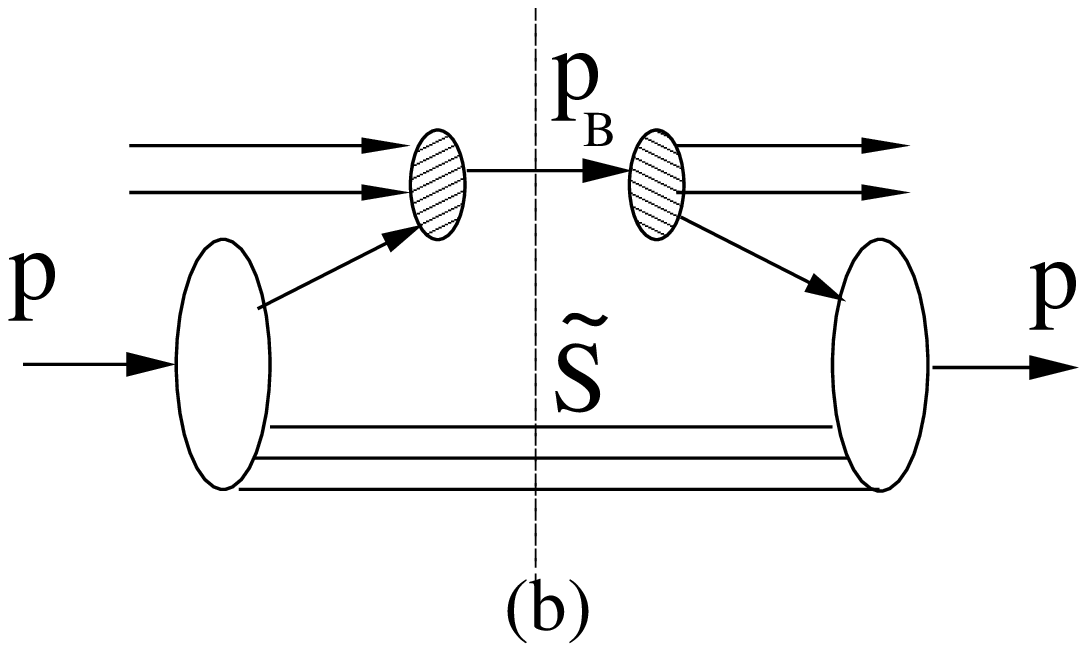}}
\resizebox{2.5in}{1.5in}{\includegraphics{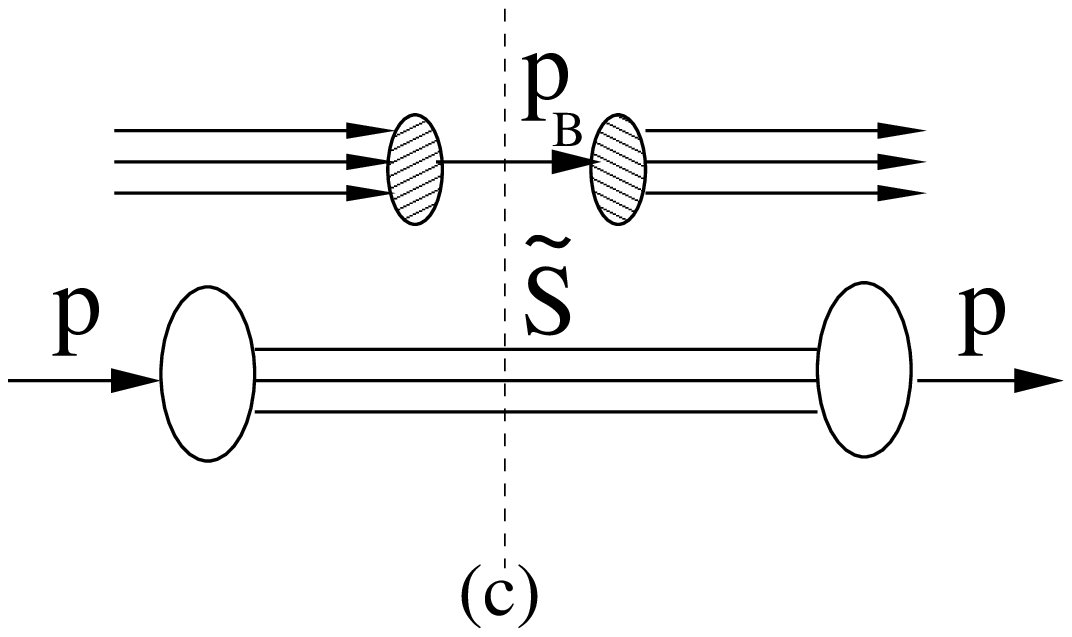}}
 \caption{The cut diagrams for contributions from (a) shower-shower-thermal,
(b) shower-thermal-thermal and (c) thermal-thermal-thermal quark
recombination to the single baryon fragmentation function.}
 \label{fig5}
 \end{figure}

The contributions from ``shower-shower-shower'' quark recombination,
denoted as ${\tilde D}_q^{B(SSS)}$, has the same expression as in
vacuum [Eq.(\ref{qfragm05})] with the replacement of the triple
quark distribution functions $F_q^{q_1q_2q_3}(z_1,z_2,z_3)$
in vacuum by its counterpart in medium ${\tilde F}_q^{q_1q_2q_3}(z_1,z_2,z_3)$.
Working along the same approach, one can also find out
the contribution from ``thermal-thermal-thermal'' quark recombination,
 \begin{eqnarray}
   {\tilde D}_{q}^{B(TTT)}(z_B, p^+)
   &=&\int\frac{d^2q_{1\perp}dz_{q_1}}{2(2\pi)^3}
   \int\frac{d^2q_{2\perp}dz_{q_2}}{2(2\pi)^3}
   \int\frac{d^2q_{3\perp}dz_{q_3}}{2(2\pi)^3}
   f_{q_1}(q_{1\perp}, z_{q_1}p^+)f_{q_2}(q_{2\perp}, z_{q_2}p^+)
   f_{q_3}(q_{3\perp}, z_{q_3}p^+)S_{q}^0(z_B)
 \nonumber\\
   & & \hspace{-0.5in}\times\frac{4}{z_B}\left|\varphi_B (q_{1\perp}, \frac{z_{q_1}}{z_B}; q_{2\perp},
   \frac{z_{q_2}}{z_B}; q_{3\perp},\frac{z_{q_3}}{z_B})\right|^2
   \left[(2\pi)^3\delta^{(2)}(p_{B\perp}{-}q_{1\perp}{-}q_{2\perp}{-}q_{3\perp})
   \delta(z_B{-}z_{q_1}{-}z_{q_2}{-}z_{q_3})\right]^2
   \, ,
 \label{baryonTT1}
 \end{eqnarray}
where $S_{q}^0(z_B)$ is the same factor associated with phase-space
integration as in Eq.~(\ref{dq0}). After extracting the spatial volume
from the $\delta$-function squared, one has
 \begin{eqnarray}
   {\tilde D}_{q}^{B(TTT)}(z_B, p^+)
   &=& V p^+ \int\frac{d^2p_{B\perp}}{(2\pi)^3}
   \int_0^{z_B}\frac{dz_{q_1}}{2z_B}\int_0^{z_B-z_{q_1}}\frac{dz_{q_2}}{2z_B}
   \int\frac{d^2q_{1\perp}}{(2\pi)^3}\int\frac{d^2q_{2\perp}}{(2\pi)^3}
 \nonumber\\
   &&f_{q_1}(q_{1\perp}, z_{q_1}p^+)f_{q_2}(q_{2\perp}, z_{q_2}p^+)
     f_{q_3}(p_{B\perp}{-}q_{1\perp}{-}q_{2\perp}, (z_B{-}z_{q_1}{-}z_{q_2})p^+)
 \nonumber\\
   &&R_B(q_{1\perp}, \frac{z_{q_1}}{z_B}; q_{2\perp}, \frac{z_{q_2}}{z_B}) \, ,
 \label{baryonTT2}
 \end{eqnarray}
which corresponds to an invariant inclusive baryon spectrum from thermal
quark recombination,
 \begin{eqnarray}
   (2\pi)^3\frac{dN^{B(TTT)}}{dp_B^+ d^2p_{B\perp}}
   &=&V\int_0^1 dx_{q_1}\int_0^{1-x_{q_1}}dx_{q_2}
   \int\frac{d^2q_{1\perp}}{2(2\pi)^3}\int\frac{d^2q_{2\perp}}{2(2\pi)^3}
   f_{q_1}(q_{1\perp}, x_{q_1}p_b^+)f_{q_2}(q_{2\perp}, x_{q_2}p_b^+)
 \nonumber\\
   &\times&f_{q_3}(p_{B\perp}{-}q_{1\perp}{-}q_{2\perp}, (1{-}x_{q_1}{-}x_{q_2})p_B^+)
   R_B(q_{1\perp}, x_{q_1}; q_{2\perp}, x_{q_2})\, .
 \label{baryonTT4}
 \end{eqnarray}

The derivation of the contribution from the ``shower-thermal-thermal'' quark
recombination to the baryon fragmentation function is also straightforward.
One has,
 \begin{eqnarray}
   {\tilde D}_{q}^{B(STT)}(z_B, p^+)
    &=&\sum_{q_3\in B}
    \int\frac{d^2 q_{1\perp}}{(2\pi)^3}\int\frac{d^2 q_{2\perp}}{(2\pi)^3}
    \int_0^{z_B}\frac{dz_{q_1}}{2z_B}
    \int_0^{z_B-z_{q_1}}\frac{dz_{q_2}}{2z_B}
    f_{q_1}(q_{1\perp}, z_{q_1}p^+)f_{q_2}(q_{2\perp}, z_{q_2}p^+)
  \nonumber\\
    &&\frac{1}{(1-z_{q_1}/z_B-z_{q_2}/z_B)^2}
    R_B(q_{1\perp}, \frac{z_{q_1}}{z_B}; q_{2\perp}, \frac{z_{q_2}}{z_B})
    {\tilde F}_q^{q_3}(z_B{-}z_{q_1}{-}z_{q_2})\, ,
 \label{bayonST12}
 \end{eqnarray}
where the summation is over different quark flavors of $q_3\in B$ among
the constituent quarks of the baryon $B$.

In deriving the contribution from ``shower-shower-thermal'' quark
recombination, one encounters again the problem of interference
between amplitudes of different shower quark recombination. We
have to make the same approximation that one can neglect the 
interference term and the final result is
proportional to the diagonal contribution with a constant $C_{B2}$.
Therefore, we have
 \begin{eqnarray}
    {\tilde D}_{q}^{B(SST)}(z_B, p^+)
    &=&\sum_{(q_1q_2)\in B}\frac{z_{B}^3}{2}\int\frac{d^4p}{(2\pi)^4}
    \delta\left(z_{B} {-} \frac{p_{B}^+}{p^+}\right)
    \int d^4 xe^{-ip\cdot x}
    \int\frac{d^2 k_{1\perp}}{(2\pi)^3}\frac{dx_{k_1}}{2\sqrt{x_{k_1}}}
    \int\frac{d^2 k'_{1\perp}}{(2\pi)^3}\frac{dx'_{k_1}}{2\sqrt{x'_{k_1}}}
    \int\frac{d^2 q_\perp}{(2\pi)^3}\frac{dx_q}{2}
  \nonumber\\
    &&\frac{1}{\sqrt{x_{k_2} x'_{k_2}}}f_{q_3}(q_{\perp}, x_{q}p_B^+)
    \varphi_B(q_{\perp}, x_{q}; k_{1\perp}, x_{k_1};k_{2\perp}, x_{k_2})
    \varphi^*_B(q_{\perp}, x_{q}; k'_{1\perp}, x'_{k_1};k'_{2\perp}, x'_{k_2})
  \nonumber\\
   &&{\rm Tr}\left\lbrack\frac{\gamma^+}{2 p_{B}^+}
    \sum_S\, _k\langle\langle {\psi}(0)
    |S; k_{1\perp}, x_{k_1}; k_{2\perp}, x_{k_2}\rangle
    \langle k'_{2\perp}, x'_{k_2}; k'_{1\perp}, x'_{k_1}; S
    |\overline\psi(x)\rangle\rangle_k\right\rbrack
  \nonumber\\
   &\approx &\widetilde{C}_{B2}\frac{z_{B}^3}{2}\sum_{(q_1q_2)\in B}\int\frac{d^4p}{(2\pi)^4}
    \delta\left(z_{B} {-} \frac{p_{B}^+}{p^+}\right)
    \int d^4 xe^{-ip\cdot x}
    \int\frac{d^2 k_{1\perp}}{(2\pi)^3}\frac{dx_{k_1}}{4x_{k_1}}
    \int\frac{d^2 q_{\perp}}{(2\pi)^3}\frac{dx_{q}}{2}
  \nonumber\\
   &&\frac{1}{x_{k_2}}f_{q_3}(q_{\perp}, x_{q}p_B^+)
    |\varphi_B(q_{\perp}, x_q; k_{1\perp}, x_{k_1};k_{2\perp}, x_{k_2})|^2
  \nonumber\\
    &&{\rm Tr}\left\lbrack\frac{\gamma^+}{2 p_{B}^+}
    \sum_S \, _k\langle\langle {\psi}(0)
    |S; k_{1\perp}, x_{k_1}; k_{2\perp}, x_{k_2}\rangle
    \langle k_{2\perp}, x_{k_2}; k_{1\perp}, x_{k_1}; S
    |\overline\psi(x)\rangle\rangle_k\right\rbrack\, ,
 \label{bayonST1}
 \end{eqnarray}
where $k_{2\perp}=p_{B\perp}{-}q_{1\perp}{-}k_{1\perp}$,
$k'_{2\perp}=p_{B\perp}{-}q_{1\perp}{-}k'_{1\perp}$,
$x_{k_2}=1{-}x_{q_1}{-}x_{k_1}$,
$x'_{k_2}=1{-}x_{q_1}{-}x'_{k_1}$, $\widetilde{C}_{B2}$ is a constant with
the dimension of momentum and the summation is over all possible
$(q_1q_2)\in B$ quark pairs among the three constituent quarks of the
baryon $B$. Changing the variables
$x_{k_1},x_{k_2},x_{q_1}$ to $z_{k_1},z_{k_2},z_{q_1}$
and neglecting the intrinsic momentum in the baryon wavefunction,
$|\varphi_B(q_{\perp}, x_{q}; k_{1\perp}, x_{k_1}; k_{2\perp}, x_{k_2})|^2
\approx R_B(q_{\perp}, x_{q}; 0_{\perp}, x_{k_1})$, we have
 \begin{eqnarray}
    {\tilde D}_{q}^{B(SST)}(z_B, p^+)
    &\approx&\sum_{(q_1q_2)\in B} \widetilde{C}_{B2}
    \int\frac{d^2 q_{\perp}}{(2\pi)^3}\int_0^{z_B}\frac{dz_{q}}{2z_B}
    \int_0^{z_B-z_q}\frac{dz}{2z_B}f_{q_3}(q_{\perp}, z_{q}p^+)
  \nonumber\\
    &\times&\frac{1}{(1-z_q/z_B)^2}
    R_B(q_{\perp}, \frac{z_q}{z_B}; 0_{\perp}, \frac{z}{z_B})
    {\tilde F}_q^{q_1 q_2}(z, z_B-z_q-z)\, .
 \label{bayonST11}
 \end{eqnarray}
The diquark distribution ${\tilde F}_q^{q_1 q_2}(z_1,z_2)$ is the same
as in the ``shower-shower'' contribution to the meson fragmentation
in Eq.~(\ref{qfragm02}), which is similarly defined as in the vacuum except
that the thermal average in principle should include medium effects such as
induced radiation and absorption.

Finally, the effective baryon fragmentation function is the sum over the
contributions resulting from ``shower-shower-shower'',
``shower-shower-thermal'', ``shower-thermal-thermal'' and 
``thermal-thermal-thermal'' constituent quark recombination,
 \begin{eqnarray}
   {\tilde D}_{q}^{B}(z_B, p^+)={\tilde D}_q^{B(SSS)}(z_B)
   +{\tilde D}_{q}^{B(SST)}(z_B, p^+)+{\tilde D}_{q}^{B(STT)}(z_B, p^+)
   +{\tilde D}_{q}^{B(TTT)}(z_B, p^+)\, ,
 \label{baryonF}
 \end{eqnarray}
though the ``thermal-thermal-thermal'' contribution is not correlated with
the initial parton jet and thus is only the background from the hadronization
of the thermal medium.

\end{widetext}

\section{Conclusion}
\label{sec6}

In this paper, we have studied medium modification to the effective
parton fragmentation functions due to quark recombination during the
hadronization of the parton jet together with a quark-gluon plasma.

We started with a formulation of the vacuum parton fragmentation functions
in the parton operator form within a constituent quark model. In this model,
constituent quarks are the effective degrees of freedom during the
hadronization of both the parton jet and the quark-gluon plasma, which
is essentially a thermalized gas of constituent quarks. Final
hadrons are composed of constituent quarks with given wavefunctions.
We showed that for sharply peaked hadron wavefunctions, one can neglect
the interference between amplitudes of hadron formation from
recombination of constituent quarks with different momenta.
Consequently, we were able to cast the meson (baryon) fragmentation 
functions into a convolution of the recombination probability and 
constituent diquark (triquark) distribution functions of the 
fragmenting parton. The recombination probability is
determined by the hadron's wavefunction in the constituent quark model.
The diquark or triquark quark distribution functions of the fragmenting
parton are defined as the overlapping matrices between parton field
operator and the final constituent quark states, similarly as the
dihadron or trihadron fragmentation functions defined as the overlapping
matrices between parton operator and final hadron states. We derived
the QCD DGLAP evolution equations for the diquark and triquark
distributions functions which are a little different from that of
dihadron and trihadron fragmentation functions because of the
kinematic constraints imposed by the hadrons' wavefunctions in the
constituent quark model. We further discussed possible connections
between triquark, diquark and single quark distribution functions
through sum rules.

Working within the framework of field theory at finite temperature,
we extended the formulation of parton fragmentation functions in
terms of quark recombination to the case of jet fragmentation in
medium, assuming that the medium is hadronizing together with the parton
jets. Replacing the vacuum expectation in the overlapping matrices
by the thermal average, we were able to derive the medium
modification to the fragmentation functions via quark recombination.
The medium modification comes not only as the medium modification
of the diquark or triquark distribution functions in the
recombination of constituent quarks from the fragmenting parton,
called shower quark recombination, but also as additional
contributions from recombination between shower and thermal
constituent quarks.
We also obtained contributions from
recombination of thermal quarks within the same formalism.
However, such contributions of thermal quark recombination
are disconnected with parton jet fragmentation and therefore
are just a thermal background to jet fragmentation processes
in the medium.

The formulation and derivation of jet fragmentation functions in terms 
of quark recombination in the vacuum do not simplify the theoretical
description of the jet fragmentation processes. It is only a
model of the non-perturbative process of hadronization. However,
once extended to jet fragmentation in medium, it provides a
natural framework for the description of jet hadronization
in medium and the study of effective medium modification to the
fragmentation functions. Within this formalism, one can not only
include contributions to jet fragmentation from the recombination
of constituent quarks from the parton jet and the thermal medium but also
the medium modification of the quark distributions of the jet parton
due to induced gluon bremsstrahlung and absorption.
This framework will effectively unify parton energy loss via
induced gluon radiation and quark recombination processes during
hadronization.


\acknowledgements

We would like to thank R. Hwa for stimulating discussion and
helpful comments on the manuscript. This work was supported by
by the Director, Office of Energy Research, Office of High Energy and
Nuclear Physics, Division of Nuclear Physics, and by the Office of
Basic Energy Science, Division of Nuclear Science, of the U.S.
Department of Energy under Contract No. DE-AC03-76SF00098, and 
by National Natural Science Foundation of China under projects
10440420018, 10475031 and 10135030, and by MOE of China under
projects NCET-04-0744, SRFDP-20040511005, CFKSTIP-704035. 
E.~W. thanks LBNL Nuclear Theory Group for its hospitality during the
completion of this work.


\end{document}